\documentclass[preprint,review,3p,12pt,authoryear,sort&compress]{elsarticle}
\usepackage[colorlinks=true, 
                      citecolor=blue,
                      linkcolor=blue,
                      urlcolor=blue
                       ]{hyperref} 
  
\usepackage{natbib}
\usepackage{lineno}
\modulolinenumbers[5]
\usepackage{amssymb}
\usepackage[applemac]{inputenc}
\usepackage{amsmath}

\usepackage{enumerate}

\journal{International Journal of Plasticity}

\biboptions{authoryear}

\begin{document}

\begin{frontmatter}



\title{Transformation-mediated Plasticity in CuZr based Metallic Glass Composites: A Quantitative Mechanistic Understanding}

\author[a,b]{B.A.~Sun}\corref{cor1}
\ead{iphysunba@gmail.com}
\author[c]{K.K.~Song}
\author[c]{S.~Pauly}
\author[d]{P.Gargarella}
\author[b]{C.T.~Liu}
\author[e,f]{J.~Eckert}
\author[b]{Y.~Yang\corref{cor1}}
\ead{yonyang@cityu.edu.hk}
\cortext[cor1]{Corresponding author}
\address[a]{Materials Genome Institute, Shanghai University, Shanghai 200444, China}
\address[b]{Centre for Advanced Structural Materials, Department of Mechanical and Biomedical Engineering, City University of Hong Kong, Kowloon, Hong Kong}
\address[c]{IFW Dresden, Institut f\"ur Komplexe Materialien, Helmholtzstra\ss e 20, D-01069, Dresden,Germany}
\address[d]{Departamento de Engenharia de Materiais, Universidade Federalde S\~ao Carlos,13565-905 S\~ao Carlos, S\~ao Paulo, Brazil}
\address[e]{Erich Schmid Institute of Materials Science, Austrian Academy of Sciences, Jahnstra\ss e 12, A-8700 Leoben, Austria}
\address[f]{Department Materials Physics, Montanuniversit\"at Leoben, Jahnstra\ss e 12, A-8700 Leoben, Austria}%





\begin{abstract}
In this paper, we present a thorough stress analysis of the Cu-Zr metallic-glass composite with  embedded B2 particles subject to a martensitic transformation. Within the framework of the Eshelby theory, we are able to explain, in a quantitative manner, (1) the formation of three types of shear bands with distinct morphologies as observed experimentally in the severely deformed Cu-Zr metallic-glass composite and (2) the work hardening ability of the Cu-Zr metallic-glass composite as related to the coupled effects of elastic back stress and elastic mismatch caused by the martensitic transformation. Furthermore, we also discuss the issues about the stress affected zone of the individual B2 phase and the stability of the crystalline-amorphous interface. Given the general agreement between the theoretical and experimental findings, we believe that the outcome of our current work can lead to a deeper understanding of the transformation-induced plasticity in the Cu-Zr based metallic glass composites, which should be very useful to the design of the metallic-glass composites with improved ductility. 
\end{abstract}

\begin{keyword}
Metallic glass \sep A.ductility \sep A.phase transformation \sep C.analytic functions
\end{keyword}

\end{frontmatter}


\section{Introduction}


Bulk metallic glasses (BMGs)\citep{Wang200445,Chen:2011aa} are a new class of advanced materials, which have attractive mechanical properties such as high strengths and large elastic limits arising from their long-range disorder atomic structure. Nevertheless, their applicability as structural materials is largely suppressed by the poor ductility\citep{Schuh20074067,Ashby2006321,Greer201371,Fornell20091540,Zhou2013147}, which has been regarded as the Achill's heel of BMGs. At room temperature, plasticity in BMGs is an inhomogeneous process with plastic strain highly localized into shear bands\citep{Greer201371,Jiang20081,Chen20101645,Chen201318,Huang201487}. Once initiated, shear bands are prone to becoming unstable due to strain softening, causing the catastrophic failure of a BMG sample along the main shear plane\citep{Shimizu20064293,Sun2015211,Wu2015136, Chen201654}. Consequently, this results in almost zero tensile ductility or very limited compressive plasticity at the macroscopic scale\citep{Zhang20031167,PhysRevLett.110.225501,Wu2011560}. In the past decades, tremendous efforts\citep{Lee20044121,PhysRevLett.84.2901,Hofmann:2008aa,Hofmann23122008,ChoiYim19992455,Liu1385,Yu2009640,Eckert1998595,Lewandowski2005,Jang2011858,Yoo2012108,Tong2016141}have been made to overcome the intrinsic weakness of BMG, among which one effective approach is to fabricate BMG matrix composites\citep{Lee20044121,ChoiYim19992455,Hofmann23122008}. This often involves ex-situ or in-situ introduction of secondary crystalline phases into a glassy matrix. Based on this approach, significant tensile ductility was achieved in some Zr-\citep{Hofmann:2008aa} and Ti-based BMG\citep{Hofmann23122008} composites. However, in the early development, the BMG composites still displayed macroscopic strain softening under tension\citep{Hofmann:2008aa}, which led to necking instability upon yielding, a phenomenon that is not desirable in engineering applications.

Recently, CuZr-based BMG composites\citep{ADMA:ADMA201000482,Pauly:2010aa,Hofmann10092010,Wu20112928,Song2012132,Liu20123128,Gargarella2014259} have attracted a great deal of research interest because of their unique mechanical properties. Unlike other types of BMG composites, these CuZr-based BMG composites comprise shape-memory B2 CuZr crystalline phases, which could undergo martensitic transformation (MT) during deformation and turn into the B19' phase. By a careful control of the size and distribution of the B2 phases, it has been shown\citep{ADMA:ADMA201000482, Pauly20104883} that these composites can exhibit significant tensile ductility and also display pronounced work-hardening, an indispensable ''ingredient'' for the practical application of structural materials. Apparently, these superior mechanical properties are derived from the interaction between the MT of B2 phases and the plastic flow ''carriers'', i.e. shear bands, in the glassy matrix. Furthermore, it was proposed that the transformation of B2 might exert a compressive stress on the glassy matrix and thus retards the rapid propagation of shear bands and/or cracks\citep{Pauly20104883,Hofmann10092010}. To validate this proposition, finite element (FE) simulations were utilized to obtain the stress/strain fields in the deformed BMG composites using the commercial packages\citep{Wu20112928,Gargarella2014259,Liu:2014aa}; however, these preliminary analyses are generally insufficient because the transient phase transformation behavior of the B2 phase is not considered in the constitutive model of the commercial FE software\citep{Leblond1989551}. Therefore, to understand the mechanisms underlying the tensile plasticity in the CuZr-based BMG composites, a precise knowledge of the stress redistribution around the transforming B2 phases is still lacking. 

In this study, we develop a micromechanical model based on the Eshelby theory\citep{Eshelby,WeibergerC} to understand the plasticity associated with the MT of B2 in the CuZr BMG composites.  It will be shown that the model predictions, such as the stress state and the resultant distribution of shear bands and strain hardening coefficient, compare very well with the experimental observations. Our current study provides a quantitative understanding of the mechanisms underlying the strain hardened plasticity in the CuZr-based BMG composites, which should be also helpful for a better control of the microstructure in BMG based composites with improved mechanical properties.

\section{Experiments and Results}

\subsection{Sample Preparations and Experimental Set-Up}

Master alloy ingot with a nominal composition Cu$_{48}$Zr$_{48}$Al$_{4}$ were prepared by arc melting the mixture of the constitute elements with purity $>99.5\%$ in a Ti-getted atmosphere. Each alloys ingots were remelted at least three times for the compositional homogeneity. Plate samples with a thickness of 2 mm, a width of 10 mm and  a length about 80 mm were obtained by suction casting into the copper mould. As the morphology of B2 phase is very sensitive to experimental condition in casting process such as the cooling rate and the quench temperature, we control the melting temperature by applying the arc current of 320 A on each alloy ingot for about 20 second before quenching to ensure the microstructure reproducibility of the BMG composite. The specimens with a dimension of about $2\times2\times4$ mm were cut from the lowest part of the plate (the part with the highest cooling rate) with a diamond saw, and then carefully ground into samples with an aspect ratio of 2:1 for the compression tests. The four side faces of samples are also carefully polished for the structural characterization after deformation. The uniaxial compression tests were performed with an Instron 5869 electromechanical test system with the maximum load of 50 KN at a constant strain rate of $2.5\times10^{-4} \textnormal{s}^{-1}$. The strain were measured by a laser extensometer (Fiedler) attached to the testing machine. Each compression test was repeated at least three times to ensure the reproducibility of deformation results. After deformation, the morphology of the shear bands as well as the fracture surface are investigated by the scanning electron microscopy (SEM, Gemini 1530).The structure nature of as-cast as well as post-deformed samples is examined by the X-ray diffraction (XRD, PANalytical X'Pert PRO) with Co-Ka radiation (the wave length $\lambda=1.7902$ nm) and an Zeiss Axiophot Optical Microscopy (OM). 

\subsection{Microstructures of the Cu$_{48}$Zr$_{48}$Al$_{4}$ BMG Composite}

\begin{figure}
 \centering
\includegraphics[width=100mm]{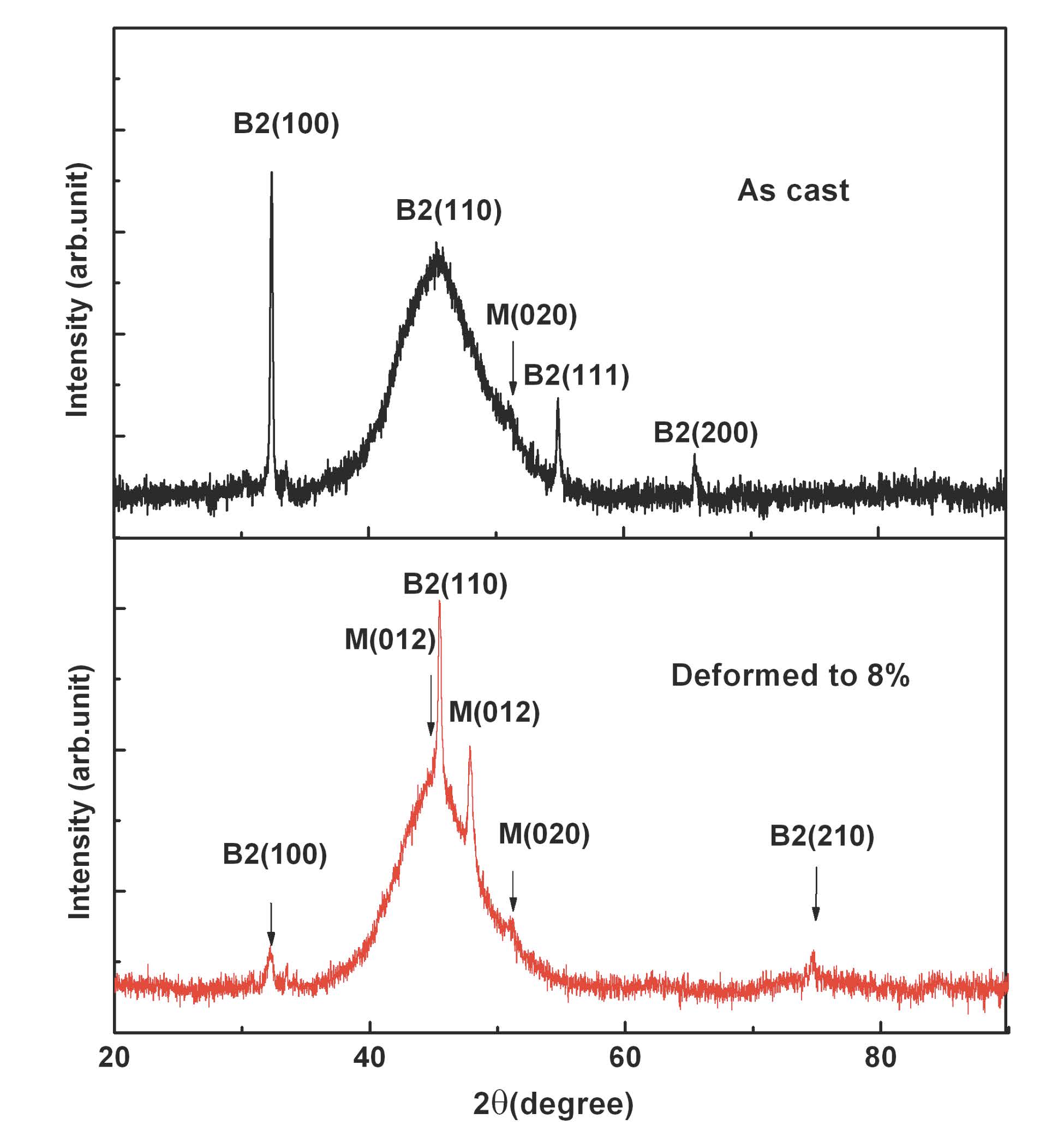}
\caption{\label{fig1} The XRD curves for the as-cast and deformed Cu$_{48}$Zr$_{48}$Al$_{4}$ (up to $\epsilon_{p}=8\%$) metallic glass composites. The B2 phase and martensite phase are identified.}
\end{figure}

At first, we need to choose a CuZr-based BMG as the base material. According to the previous study\citep{Yu20081,Qiao2016}, the glass-forming ability (GFA) and plasticity of  the Cu-Zr-Al ternary BMGs strongly depends on the content of Al. On one hand, it was reported that, when the content of Al is around 4$\sim$5\% (in atomic percent), the Cu-Zr-Al BMGs could possess a relatively large value of Poisson's ratio and show a plastic strain of ~18\% before failure\citep{Yu20081,PhysRevLett.94.205501}. On the other hand, the Cu$_{48}$Zr$_{48}$Al$_{4}$ BMG can be cast into a fully amorphous 2-mm-diameter rod or a 1-mm-thickness plate\citep{TangMB}, indicative of a very good GFA better than that of the typical binary Cu$_{50}$Zr$_{50}$ BMG. It was noticed that further addition of Al into the Cu$_{48}$Zr$_{48}$Al$_{4}$ BMG would enhance the GFA but greatly deteriorate the plasticity. Therefore, we finally chose the Cu$_{48}$Zr$_{48}$Al$_{4}$ BMG as the matrix material and cast it into a 2-mm-thick plate to obtain the BMG composite. \textcolor{blue}{Figure} \ref{fig1} shows the typical XRD pattern of the as-cast sample taken from the lowest part of the plate (the part with highest cooling rate), from which one can clearly see the crystalline peaks superimposed on the broad halo of the amorphous matrix. The majority of the crystalline peaks can be easily identified to be the B2 CuZr phase while the (001) peak of the martensitic phase B19' could be also found, suggesting a small volume fraction of the B19' phase in the as-cast sample. In Ref.34, a similar phenomenon was observed and the small amount of the B19' phase was attributed to the residual-stress induced martensitic transformation of the B2 phase during solidification\citep{Song20126000}. After the deformed to 8\%, obvious crystalline peaks of the martensite phase (B19') appeared in the XRD pattern (see \textcolor{blue}{Fig.\ref{fig1}(b)}), indicating that some B2 phases underwent a martensite phase transition during the deformation.

\begin{figure}\centering
\includegraphics[width=100mm]{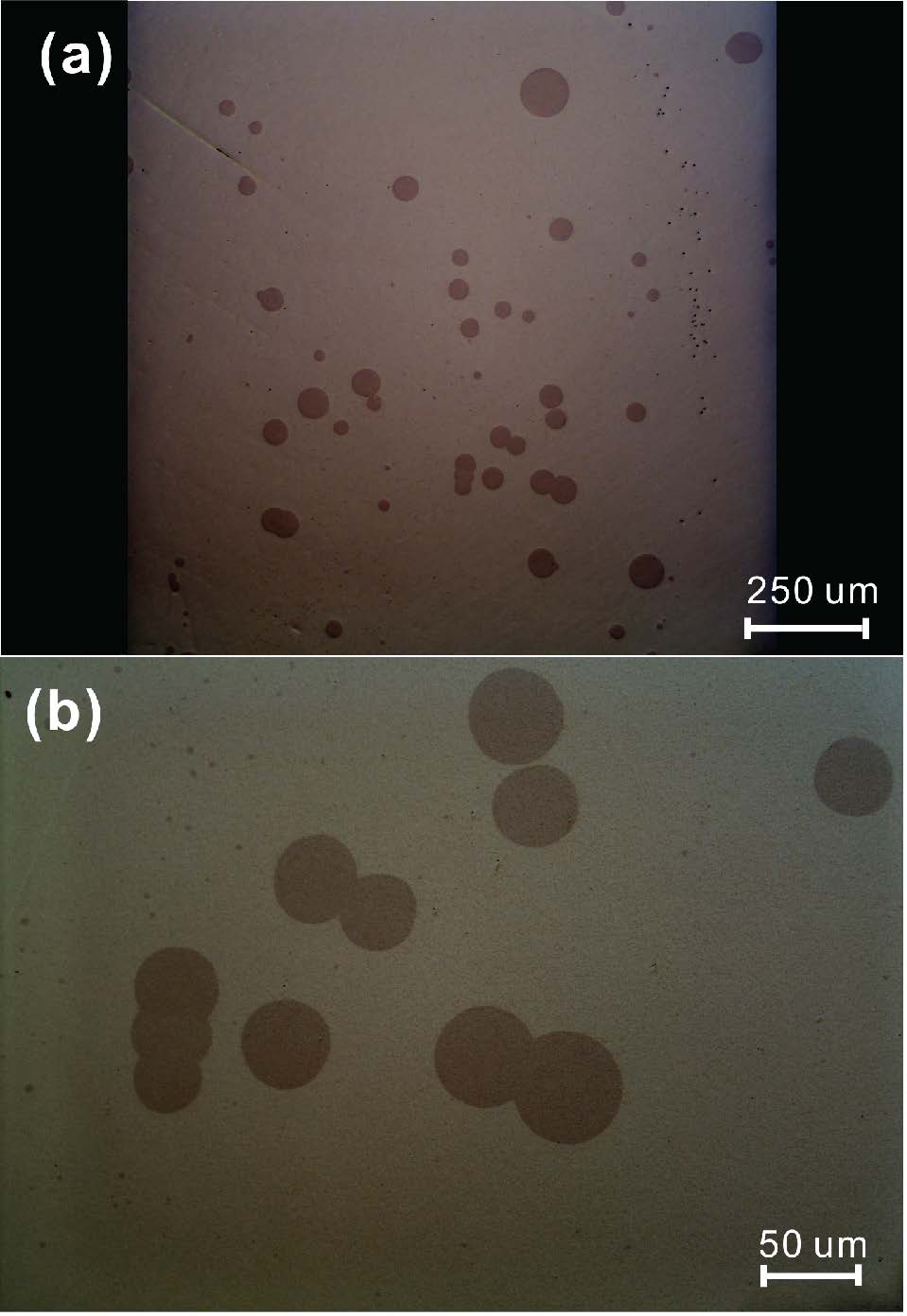}
\caption{\label{fig2} The microstructure of as-cast Cu$_{48}$Zr$_{48}$Al$_{4}$ metallic glass composite plate under optical microscope. One can see that spherical B2 phases are homogeneously distributed in the glassy matrix.}
\end{figure}

  As aforementioned, the size, morphology and distribution of the B2 CuZr phase in the glassy matrix are very sensitive to the alloy composition and experimental parameters, such as the cooling rate and the melting temperature employed for the casting process\citep{SPaulyapl,Song2012132,Wu20112928}. Furthermore, recent studies\citep{Liu20123128} also showed that a homogeneous distribution of B2 particles could be achieved by the minor addition of some alloying elements, such as Co and Ta, into the glassy matrix. \textcolor{blue}{Figure} \ref{fig2} displays the morphology and distribution of the B2 CuZr phase on the cross-sections of our as-cast samples. As seen in this figure, there is a homogeneous distribution of the spherical B2 phases in the glassy matrix, which is different from those typically reported Cu-Zr-Al composites, where the heterogeneous distribution of the B2 phases and even the dendritic structure can be found\citep{SPaulyapl,Song2012132}. As shown in \textcolor{blue}{Fig.}\ref{fig2}, the diameters of these spherical particles are in the range of 25-100 $\mu$ m with an estimated average of $\sim$42 $\mu$m. Here, it is worth noting that the spherical B2 phases are distributed in the 3D glassy matrix and therefore their volume fraction can be not simply estimated from the 2D cross sectional view, as shown in \textcolor{blue}{Fig.}\ref{fig2}. However, by assuming a thin foil projection\citep{Pauly:2010aa}, the volume fraction of the B2 phases can be roughly estimated as:  

 \begin{equation}\label{eq1}
 f_{B2}=(\frac{-2\pi \overline{D}_{B2}}{\pi\overline{D}_{B2}+8t})\ln(1-A)
 \end{equation}\\
 where $\overline{D}_{B2}$ is the mean diameter of B2 spheres (here, $\overline{D}_{B2}=42$ $\mu$m), $t$ is the thickness of the foil and is roughly equal to$\overline{D}_{B2}$, and $A$ is the project area fraction from the cross section (for example, $A\approx8\%$ for \textcolor{blue}{Fig.}\ref{fig2}\textcolor{blue}{a}). The volume fraction of B2 CuZr phase calculated by \textcolor{blue}{Eq.}\ref{eq1} is about 5-8\% for all the samples obtained via the copper-mold casting after an appropriate selection of the casting parameters.   

 \subsection{The Stress-Strain Curve of Cu$_{48}$Zr$_{48}$Al$_{4}$ BMG Composite}
 
\begin{figure}\centering
\includegraphics[width=120mm]{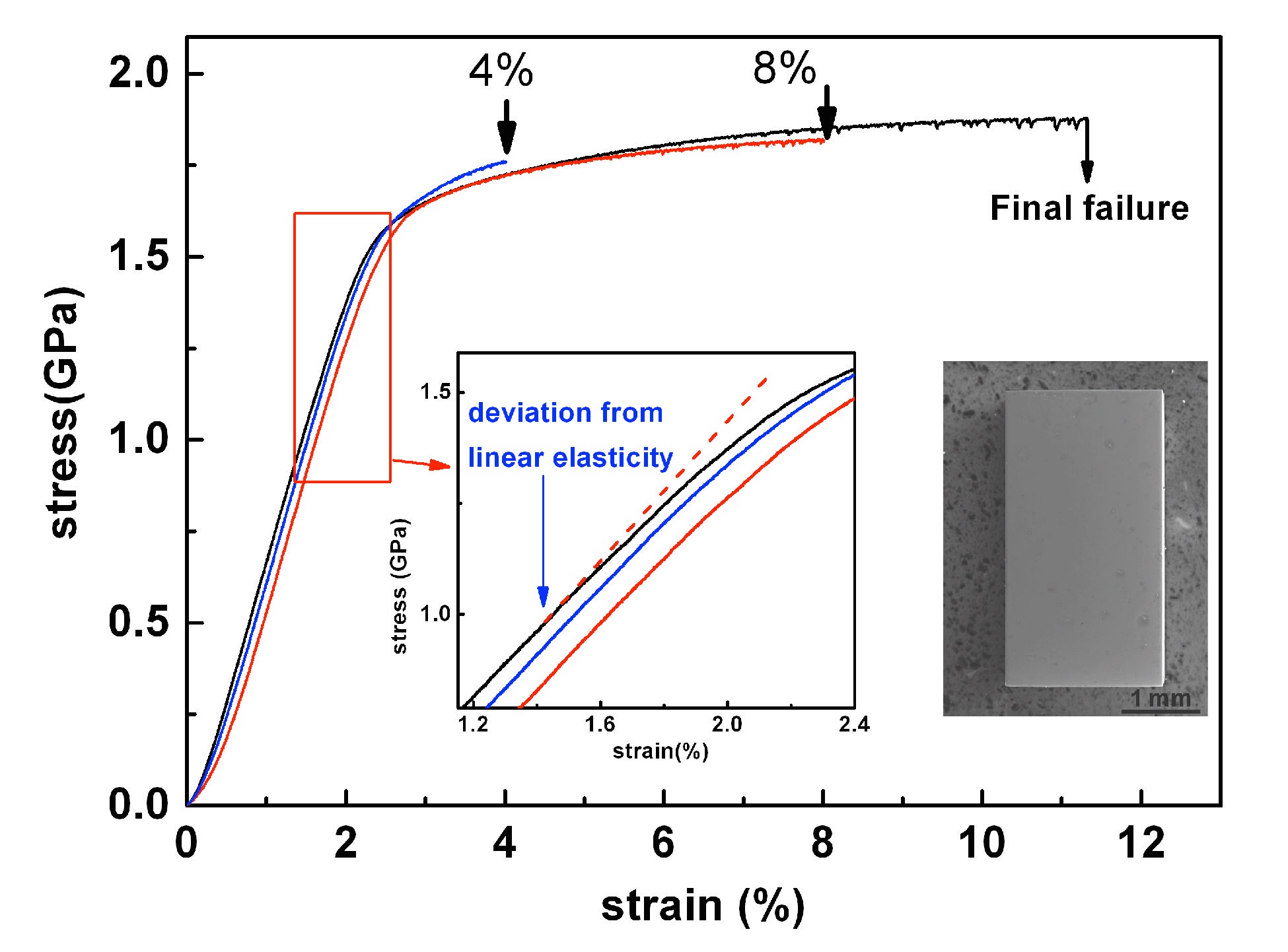}
\caption{\label{fig3} The stress-strain curves of Cu$_{48}$Zr$_{48}$Al$_{4}$ metallic glass composites deformed to different strain levels (4\%, 8\% and final fracture) in compression. The inset shows that the stress begins to deviate the linear elasticity at about 1 GPa before the nominal yielding point.}
\end{figure}

 Subsequently we performed uniaxial compression tests on the Cu$_{48}$Zr$_{48}$Al$_{4}$ composite samples. To investigate shear band evolution, the compression tests were deliberately stopped at the different strains (4\%, 8\%) for an SEM observation. \textcolor{blue}{Figure} \ref{fig3} shows the typical stress-strain curves obtained from these composites. By comparison, these curves display the similar yield strength $\sim$1.5 GPa or the elastic limit of $\sim$2\% with those of the corresponding monolithic BMGs; however, the BMG composite exhibits much better ductility with a plastic strain at least 10\% before the final failure occurs. More importantly, work hardening can be observed during the deformation of the present BMG composite although it may not be as pronounced as that reported in the previous studies[20, 35], which may be due to the small volume fraction of the B2 phases (5-8\%) in the current composite. Finally, the BMG composite sample failed by the shear fracture along two shear planes initiated from two adjacent sample sides [\textcolor{blue}{Fig.\ref{fig4}(c)}], rather than alone one single primary fracture plane as seen for most monolithic BMGs. In addition, a close examination of the stress-strain curves (see the enlarged view in the rectangular region in Fig.3) shows that the deviation of the stress-strain curves from linear elasticity occurs at $\sim$1 GPa before the nominal yielding point sets out at $\sim$1.5 GPa. This is similar to the first-yielding-point phenomenon as seen in the triple-yielding-behavior of the CuZr-based BMG composites that contain a large volume fraction of B2 phases\citep{Song20126000}. This makes sense since the activation stress for MT of B2 is known to be lower than the typical yielding strength of a glassy matrix\citep{SPaulyapl}. Therefore, the initial yielding point of $\sim$1 GPa (it makes sense to see that the yield stress of B2 is much less than BMG materials) as we herein observed strongly indicates that the MT of B2 occurs prior to the overall yielding of the composite, which causes stress redistribution in the glassy matrix and thus affects the subsequent shear banding, as will be discussed in the later text. 

\subsection{Shear-Band Morphology in BMG Composite} 
\begin{figure}\centering
\includegraphics[width=100mm]{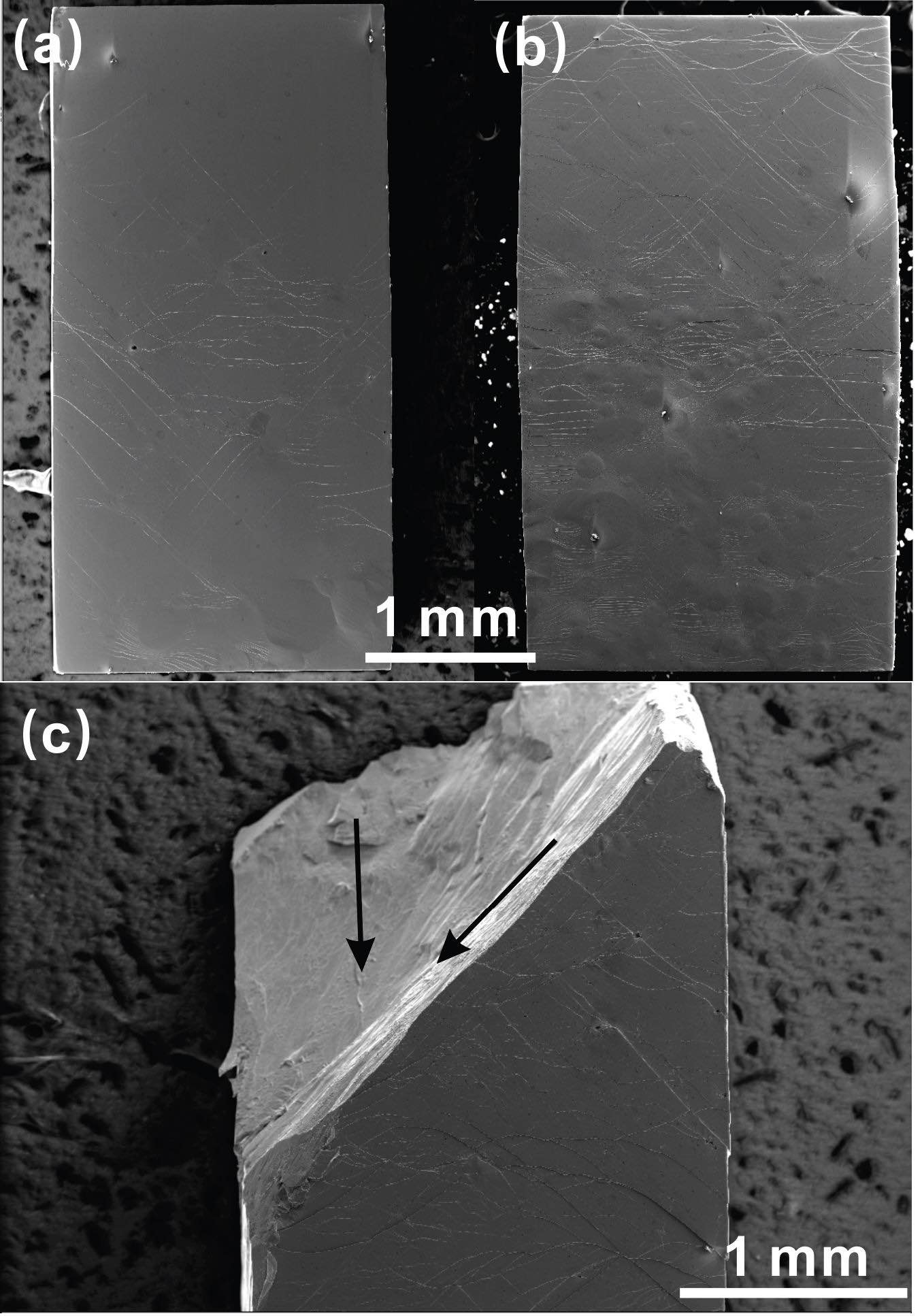}
\caption{\label{fig4} The SEM images shows the overall morphology of Cu$_{48}$Zr$_{48}$Al$_{4}$ samples after the deformation: (a) 4\%; (b) 8\% and (c) the final fracture.}
\end{figure}

\begin{figure}\centering
\includegraphics[width=150mm]{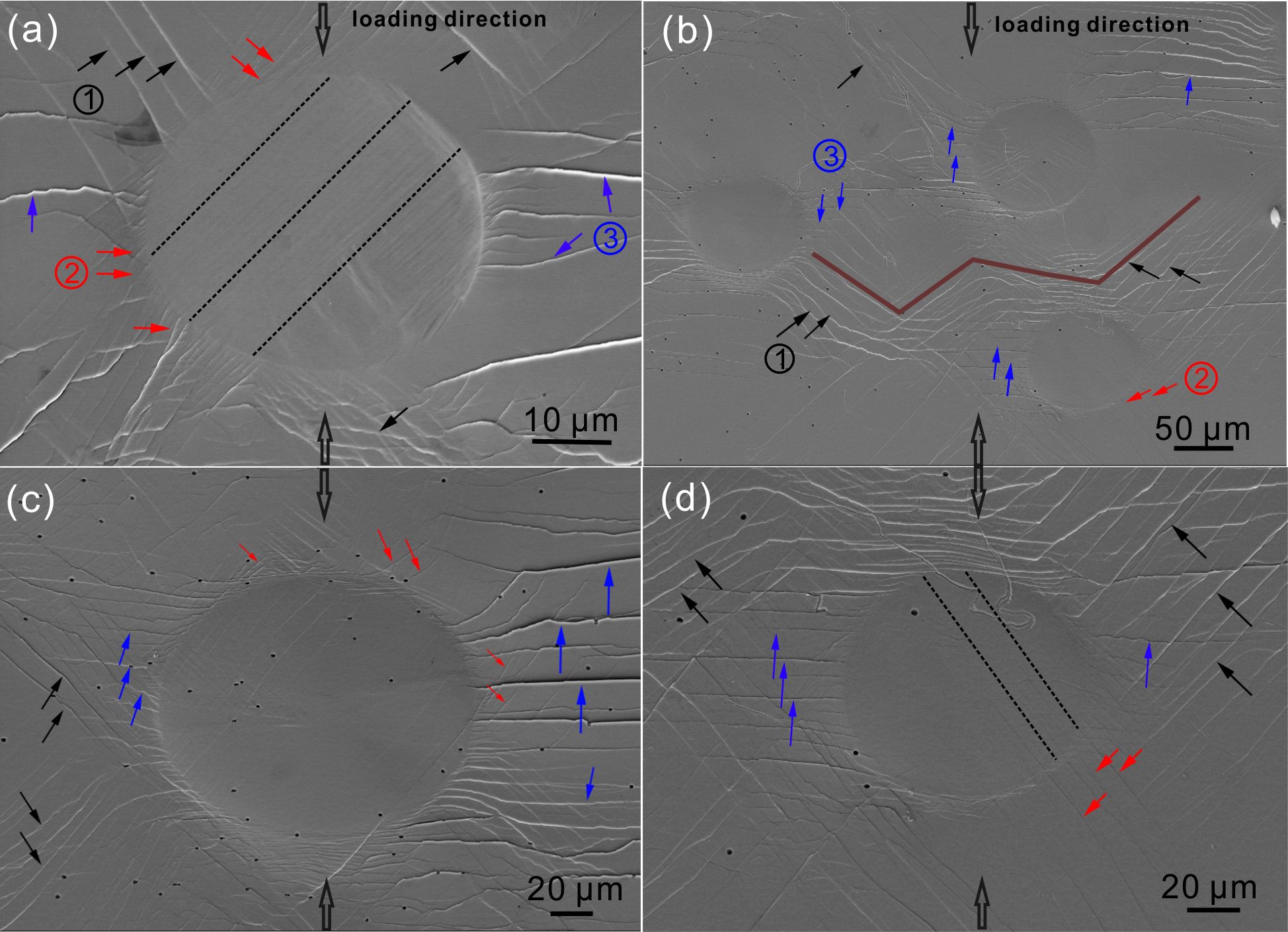}
\caption{\label{fig5} The shear-band distributions on the surface morphology of the deformed Cu$_{48}$Zr$_{48}$Al$_{4}$ samples. Three types of shear band are identified and labeled by black, red and blue arrows, respectively. }
\end{figure}

After the deformation, the surface shear-band morphology was systematically investigated under SEM. As shown in \textcolor{blue}{Fig. }\ref{fig4}, multiple shear bands can be clearly observed. Especially, many fine and wavy shear bands can be found around the crystalline phases. These shear bands are sparsely distributed at the relatively low strain$\sim$4\% and gradually become densely populated across the whole sample with the increasing strain, as shown in \textcolor{blue}{Fig. \ref{fig4}}(b) and (c), respectively. A careful examination of these SEM images reveals that these complex shear bands can be classified into three types. As indicated by the black arrows in \textcolor{blue}{Fig. \ref{fig5}}, the type-I shear bands are within the glassy matrix and away from the crystalline particles. Like the shear bands in monolithic BMGs\citep{PhysRevLett.105.035501}, the type-I shear bands are oriented along the inclined direction of about 45$^{\circ} $ to the loading direction, conforming to the direction of the maximum shear stress. Here it is worth mentioning that the type-I shear bands could be deflected and change their direction of propagation when they approach the crystalline particles, as shown in \textcolor{blue}{Fig. \ref{fig5}}(b) and (c). By comparison, the Type-II shear bands seemingly ''emanate'' from the crystalline phases and are well aligned along the direction of the martensitic transformation ''stripes''.  Compared to the type-I shear bands, these type-II shear bands are short and finely spaced, which disappear within a short distance from the glass/crystalline interface, as indicated by the red arrows in \textcolor{blue}{Fig. \ref{fig5}}. Since the B2 particles embedded in the glassy matrix have different crystal orientations, the direction of the martensitic transformation ''stripes'' also varies from one particle to another. As a result, there is no fixed direction for the type-II shear bands. Finally, it is worth pointing out that the type-III shear bands, are also attached with the crystalline phases and are oriented along the fixed direction perpendicular to the loading axis, as indicated by blue arrows. According to our SEM observations, the three types of shear bands mutually interact and intersect with each other, which contributes to the overall plasticity in the BMG composite and also provides the vivid experimental evidence to decode the underlying deformation mechanisms.  

\section{Theory}

Theoretically, the martensitic transformation of the embedded B2 phase results in a stress and strain redistribution in the glassy matrix, which may cause the deflection of a propagating shear band, such as the type-I shear band (\textcolor{blue}{Fig. \ref{fig5}(d)}), and/or even affect the initiation of an embryonic shear band, such as the type-II and the type-III shear band (Fig. 5a), during the transience of a phase transformation, However, this transient effect is difficult to simulate in the conventional FEM\citep{Leblond1989551} only dealing with quasi-static equilibrium. Unlike the previous studies\citep{Wu20112928,Liu:2014aa}, here we derive the analytical expressions for the stress/strain field arising from the martensitic transformation of the B2 phases by using the Eshelby theory\citep{Eshelby,WeibergerC}. Before proceeding, it is worth noting that, in Eshelby theory\citep{Eshelby}, a region is termed as an \textit{inclusion} if it has the same elastic constants as the ''matrix'' material while a region is termed as an \textit{inhomogeneity} if it does not. For consistency, the same terminologies are also adopted in the following analysis.

\begin{figure}\centering
\includegraphics[width=120mm]{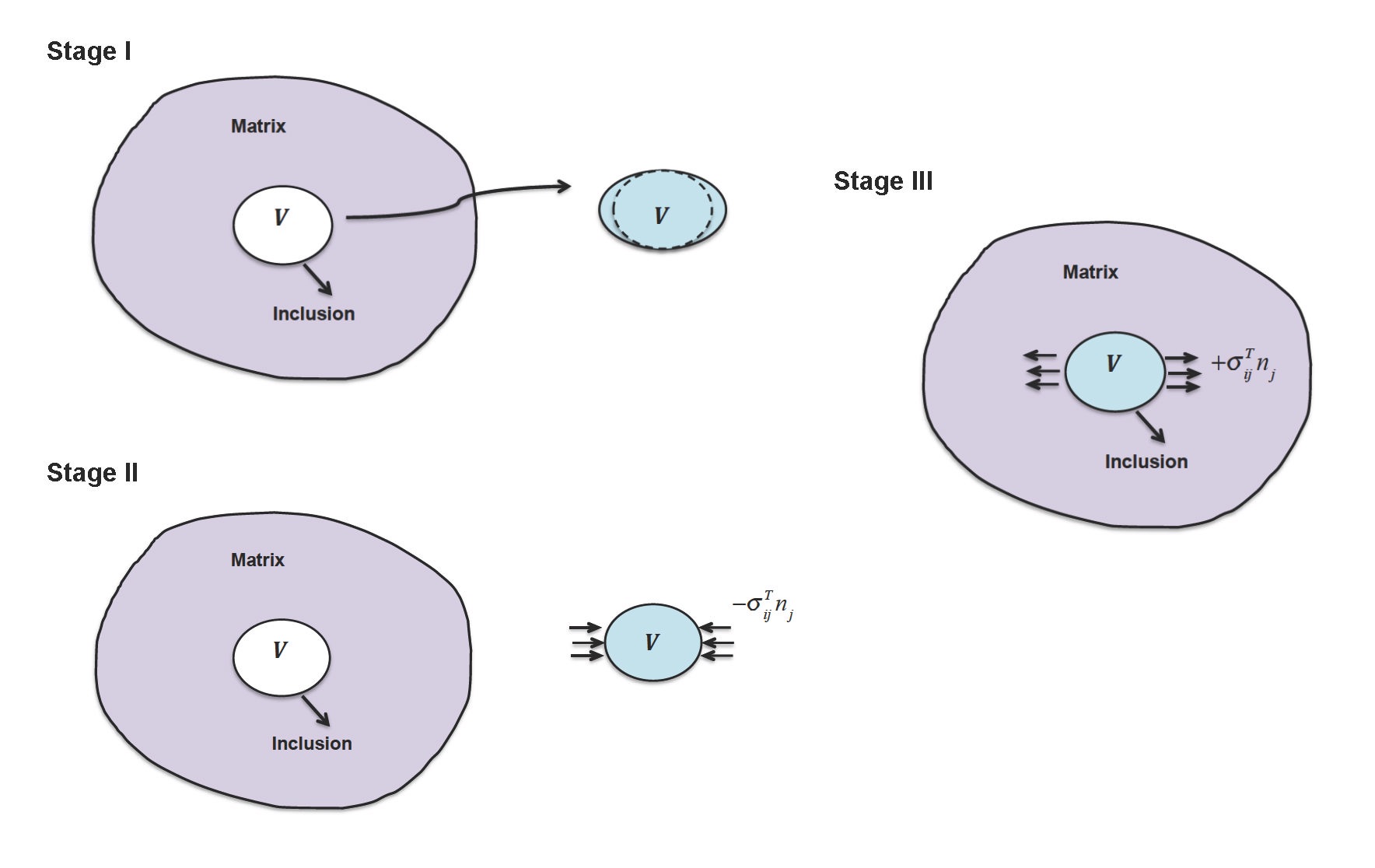}
\caption{\label{fig6} The schematic illustration of the three stages of Eshelby approach for solving the stress and strain fields due to the deformation of an inclusion in the matrix.}
\end{figure}

 In principle, a typical Eshelby problem, which entails the inelastic deformation of inclusion/inhomogeneity in an elastic matrix, can be solved following three imaginary steps\citep{Eshelby}. As shown schematically in \textcolor{blue}{Fig. \ref{fig6}}, at step I, one takes an inclusion out of the matrix and let it undergo an unconstrained transformation as characterized by the transformation strain field $\epsilon^{T}_{ij}$ ; at step II, elastic surface tractions $\sigma_{ij}^{T}$ are imposed onto the surface of the transformed inclusion such that it can fit back into the original matrix; at step III, the surface tractions are removed while the inelastic deformation remains, which leads to the stress/strain redistribution in the inclusion and elastic matrix.  Since previous experimental studies\citep{SPaulyapl} already showed that the elastic constants of the B2 phase are very close to those of the glassy matrix, the B2 phase can be here treated as an Eshelby inclusion, which is inserted in the homogeneous glassy matrix with the shear modulus $G$ and Poisson's ratio $\nu$. Following the Eshelby's approach as described above, we obtain the following expression for the elastic strain $\epsilon^{C}_{ij}$  in the matrix which is caused by the inelastic strain, occurring to a spherical inclusion with a radius $a$, during the martensitic transformation (see the detailed derivation in \ref{appendixa}), which reads: 

\begin{equation}\label{eq2}
 \begin{aligned}
 \epsilon_{ij}^{C}&=\frac{1}{2}(u_{i,j}^{C}+u_{j,i}^{C})=\frac{\sigma_{kl}^{T}}{10G(1-\nu)}
\left[ \frac{5x_{i}x_{j}x_{k}x_{l}-r^{2}(x_{k}x_{j}\delta_{il}+x_{k}x_{i}\delta_{jl})}{r^{4}} \right]+\\ \\
 &\frac{\sigma_{kl}^{T}}{6G(1-\nu)}\left(\frac{a^{3}}{r^{3}}\right) \left[(1-2\nu)+\frac{3}{5} \left(\frac{a^{2}}{r^{2}}\right)\right]
 \left[\frac{2r^{2}\delta_{il}\delta_{jk}-3(x_{k}x_{j}\delta_{il}+x_{k}x_{i}\delta_{jl})}{2r^{2}}\right]+\\ \\ 
  & \frac{\sigma_{kl}^{T}}{4G(1-\nu)}\left(1-\frac{a^{2}}{r^{2}}\right)\left(\frac{a^{3}}{r^{3}}\right)
   \left[\frac{r^{2}(x_{k}x_{l}\delta_{ij}+x_{i}x_{k}\delta_{jl}+x_{j}x_{k}\delta_{il})-10x_{i}x_{j}x_{k}x_{l}}{r^{4}}\right]
 \end{aligned}
 \end{equation}\\
where $r$ is the magnitude of the position vector for the centroid of the inclusion; $x_{i}$ is in the Cartesian coordinate ($i = 1, 2, 3$), and $\delta_{ij} = 1$ for $i = j$, and $= 0$ otherwise.  Meanwhile, the surface traction $\sigma_{ij}^{T}$: $\sigma_{ij}^{T}=\lambda\epsilon_{mm}^{T}\delta_{ij}+2\mu\epsilon_{ij}^{T}$, where $\lambda$ and $\mu$ are Lam\'e constants of the inclusion. In the inclusion, the elastic strain $\epsilon_{ij}^{C}$  is homogeneous and related with the inelastic strain  $\epsilon^{T}_{ij}$ through  $S_{ijkl}$:$\epsilon_{ij}^{C}=S_{ijkl}\epsilon_{kl}^{T}$  where $S_{ijkl}=\frac{5\nu-1}{15(1-\nu)}\delta_{ij}\delta_{kl}+\frac{4-5\nu}{15(1-\nu)}(\delta_{ik}\delta_{jl}+\delta_{il}\delta_{jk})$\citep{WeibergerC}. As a result, the additional elastic stress ($\sigma_{ij}^{C}$) in the matrix as caused by the martensitic transformation can be expressed as  $\sigma_{ij}^{C}=\lambda\epsilon_{mm}^{C}\delta_{ij}+2\mu\epsilon_{ij}^{C}$. while that ($\sigma_{ij}^{I}$) in the inclusion as  $\sigma_{ij}^{I}=\sigma_{ij}^{C}-\sigma_{ij}^{T}=\lambda(\epsilon_{mm}^{C}-\epsilon_{mm}^{T})\delta_{ij}+2\mu(\epsilon_{ij}^{C}-\epsilon_{ij}^{T})$, since there is already a stress  $-\sigma_{ij}^{T}$ in the inclusion at the end of step II . Finally, the total stress in the matrix (or inclusion) is simply the sum of $\sigma_{ij}^{C}$(or $\sigma_{ij}^{I}$) and the external applied stress $\sigma_{ij}^{A}$. Here, it should be noted that both $\sigma_{ij}^{C}$ and $\sigma_{ij}^{I}$ are short-lived or instantaneous, which appear during the transience of the martenstic transformation and vanish with the phase transformation being completed.

   After the B2 phase is transformed into the martensitic phase, elastic mismatch arises because the newly formed martensitic phase possesses elastic constants different from the glassy matrix according to the previous experiments\citep{ADMA:ADMA201000482,SPaulyapl}. Consequently, this gives rise to an elastic stress field, which may be termed as elastic constraining stress, being superimposed onto the applied stress  . According to the Eshelby theory\citep{Eshelby}, the effect of the elastic mismatch on the stress/strain distribution within the composite with an inhomogeneity is equivalent to applying a effective transformation strain $\epsilon^{T}_{ij}$ to an inclusion with the size of the inhomogeneity.  Assume that the bulk (or shear) modulus of the martensitic phase is $K_{1}$ (or $G_{1}$) while that of the glassy matrix is $K$ (or $G$), it can be derived that the effective transformation strain[$\epsilon_{ij}^{T}=(1/3)\epsilon^{T}\delta_{ij}+$$'\epsilon_{ij}^{T}$ with $\epsilon^{T}=\epsilon_{mm}$] can be expressed as:
    \begin{equation}\label{eq3}
\epsilon^{T}=A\epsilon^{A},\qquad  '\epsilon_{ij}^{T}= B'\epsilon_{ij}^{A}
\end{equation}
where  $A$ and $B$ are the two coefficients which are functions of the elastic properties of the imhomogeneity and matrix:

\begin{equation}\label{eq4}
A=\frac{K_{1}-K}{(K-K_{1})\alpha-K},\quad  B=\frac{G_{1}-G}{(G-G_{1})\beta-G} 
\end{equation} \\
with $\alpha=(1/3)(1+\nu)/(1-\nu)$ and $\beta=(2/15)(4-5\nu)/(1-\nu)$. Here the applied far-field strain $\sigma_{ij}^{A}$ can be expressed in terms of the volumetric ($\epsilon^{A}$) and deviatoric ($'\epsilon_{ij}^{A}$) component, in which $\epsilon^{A}=\sigma^{A}/3K$, $'\epsilon_{ij}^{A}=$$'\sigma_{ij}^{A}/2G$, according to the Hook's law. With the knowledge of the $\sigma_{ij}^{T}$ from \textcolor{blue}{Eqs.}\ref{eq2} and \ref{eq3}, the elastic stress field due to the inhomogeneity can be readily obtained with the same approach as described for an Eshelby inclusion. Likewise, the total stress in the matrix (or the inhomogeneity) is the sum of $\sigma_{ij}^{C}$ (or $\sigma_{ij}^{I}$) and the applied stress $\sigma_{ij}^{A}$.

\section{Analysis and Discussion}
\subsection{Shear band distribution around crystalline phases}

Based on the above analyses, we obtained the analytical expressions for the stress/strain fields in the amorphous matrix composite during the transience of the martensitic transformation and after the transformation. Next, we would study the shear band formation in the BMG composite based on these important results. For simplicity, let us neglect the subtle pressure effect\citep{Schuh20074067,Wang201570}, although there is no technical difficulty to incorporate the pressure effect into the current analysis,  and assume that shear band formation in the glassy matrix simply follows von Mises stress criterion. For the stress field $\sigma_{ij}$ ($\sigma_{ij}=(1/3)\sigma_{mm}\delta_{ij}+$$'\sigma_{ij}$), the von Mises stress is calculated according to: $\sigma^{v}=\sqrt{(3/2)s_{ij}s_{ij}}$ with $s_{ij}=$$'\sigma_{ij}$.
\begin{figure}\centering
\includegraphics[width=150mm]{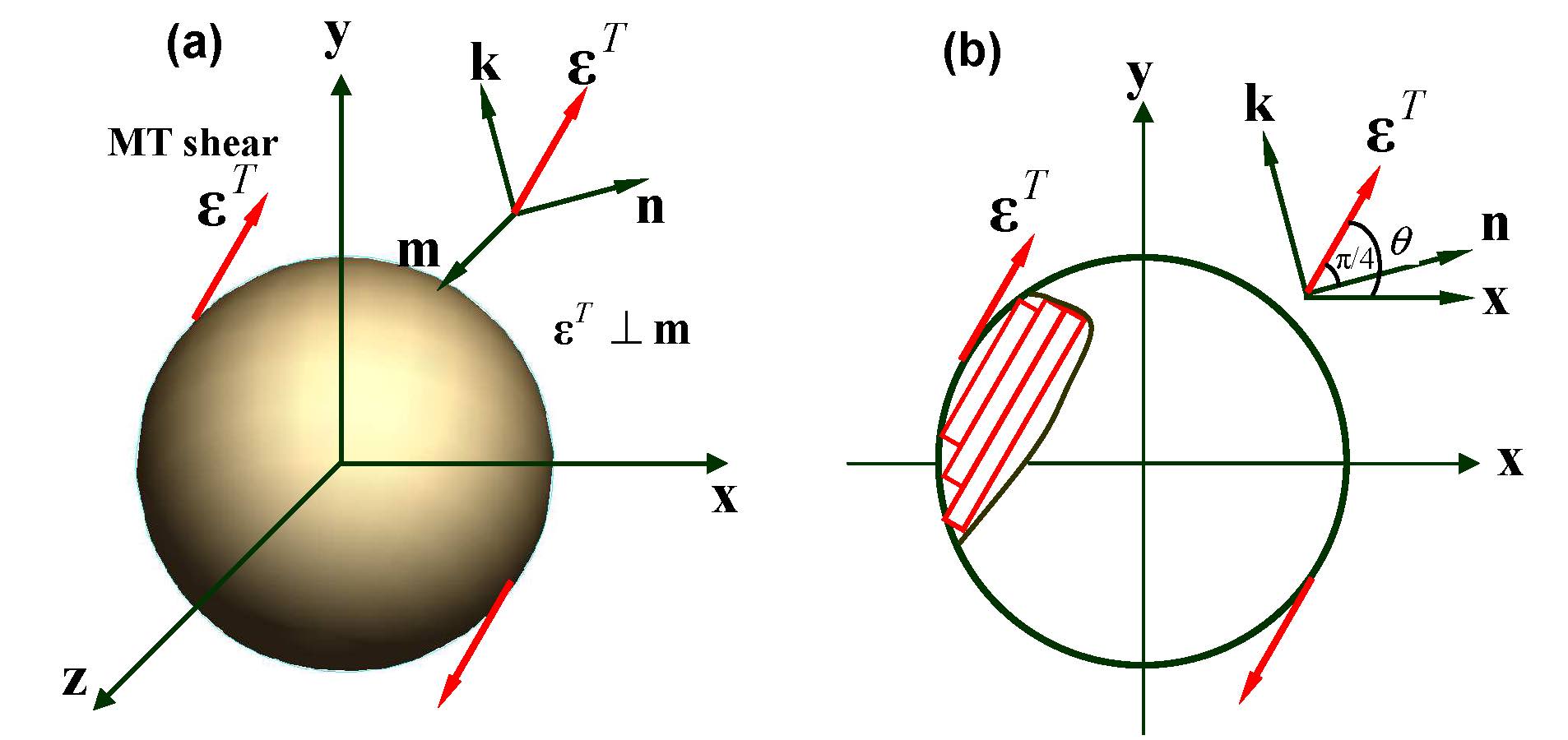}
\caption{\label{fig7} The schematic diagram illustrating the coordinate systems for the calculation of stress and strain fields and the principle axis directions of the eigenstrain due to the MT of B2 phase.}
\end{figure}

First, let us calculate the instantaneous stress field during the martensitic transformation of the B2 phases. Since the martensitic transformation is essentially shearing induced, the transformation strain (or eigenstrain) $\epsilon^{T}_{ij}$ is traceless or does not involve any volumetric change ($\epsilon^{T}_{ii}=0$). Consequently, the traceless strain tensor $\epsilon^{T}_{ij}$ can be written as $\epsilon_{ij}^{T}=2(\lambda_{n}+\lambda_{k})n_{i}n_{j}+2(\lambda_{k}+\lambda_{n})k_{i}k_{j}-(\lambda_{n}+\lambda_{k})\delta_{ij}$  where $\lambda_{n}$, $\lambda_{m}$, $\lambda_{k}$ are the eigenvalues of the deformation tensor or the three principal strains with the principal axes aligning along the unit vectors in a Cartesian coordinate system\citep{ProcacciaPRL,ProcciaPRE}. Here, we choose the coordinates so that the principal axis with the unit vector $\mathbf{m}$ coincides with the $z$ axis (see \textcolor{blue}{Fig. \ref{fig7}}). As such,  $\epsilon^{T}_{ij}$ can be simplified into the form of plane strain: $\epsilon_{ij}^{T}=\epsilon^{T}*(2n_{i}n_{j}-\delta_{ij})$, where $\epsilon^{T}$ is the magnitude of the transformation strain. Note that all components of the strain tensor $\epsilon^{T}_{ij}$ are zero except $\epsilon^{T}_{12}$ and $\epsilon^{T}_{21}$. Furthermore, we can take $n_{1}=\cos(\theta-\pi/4)$ and $n_{2}=\sin(\theta-\pi/4)$ with $\theta$ being the angle between the shear direction and the $x$ axis, as shown in \textcolor{blue}{Fig. \ref{fig6}}. In such a case, the direction of the martensitic transformation is symbolized by the angle $\theta$, which can be also used to track the direction of shear banding that is directly associated with the martensitic transformation. Once $\epsilon^{T}_{ij}$ is known, we can obtain $\sigma_{ij}^{T}$ according to $\sigma_{ij}^{T}=\lambda\epsilon^{T}_{mm}\delta_{ij}+2\mu\epsilon_{ij}^{T}$. Afterwards, we can obtain the explicit expression for $\epsilon_{ij}^{C}$ in the $x-y$ plane by substituting  $\sigma_{ij}^{T}$ into \textcolor{blue}{Eq.}\ref{eq2} (see \textcolor{blue}{Eq.}\ref{B2} in the \ref{appendixb}). For the calculation of the stress/strain field, typical values are chosen as: $\epsilon^{T}=0.01$, $G=33$ GPa, $\nu=0.373$\citep{SPaulyapl} and the far-field uniaxial compressive stress $\sigma^{A}$ is oriented along the $y$ axis. Since the martensitic transformation takes place before the overall yielding occurs in the glassy matrix, here we take $\sigma^{A}=1$ GPa according to our experimental results (\textcolor{blue}{Fig. \ref{fig3}}). 
\begin{figure}\centering
\includegraphics[width=100mm]{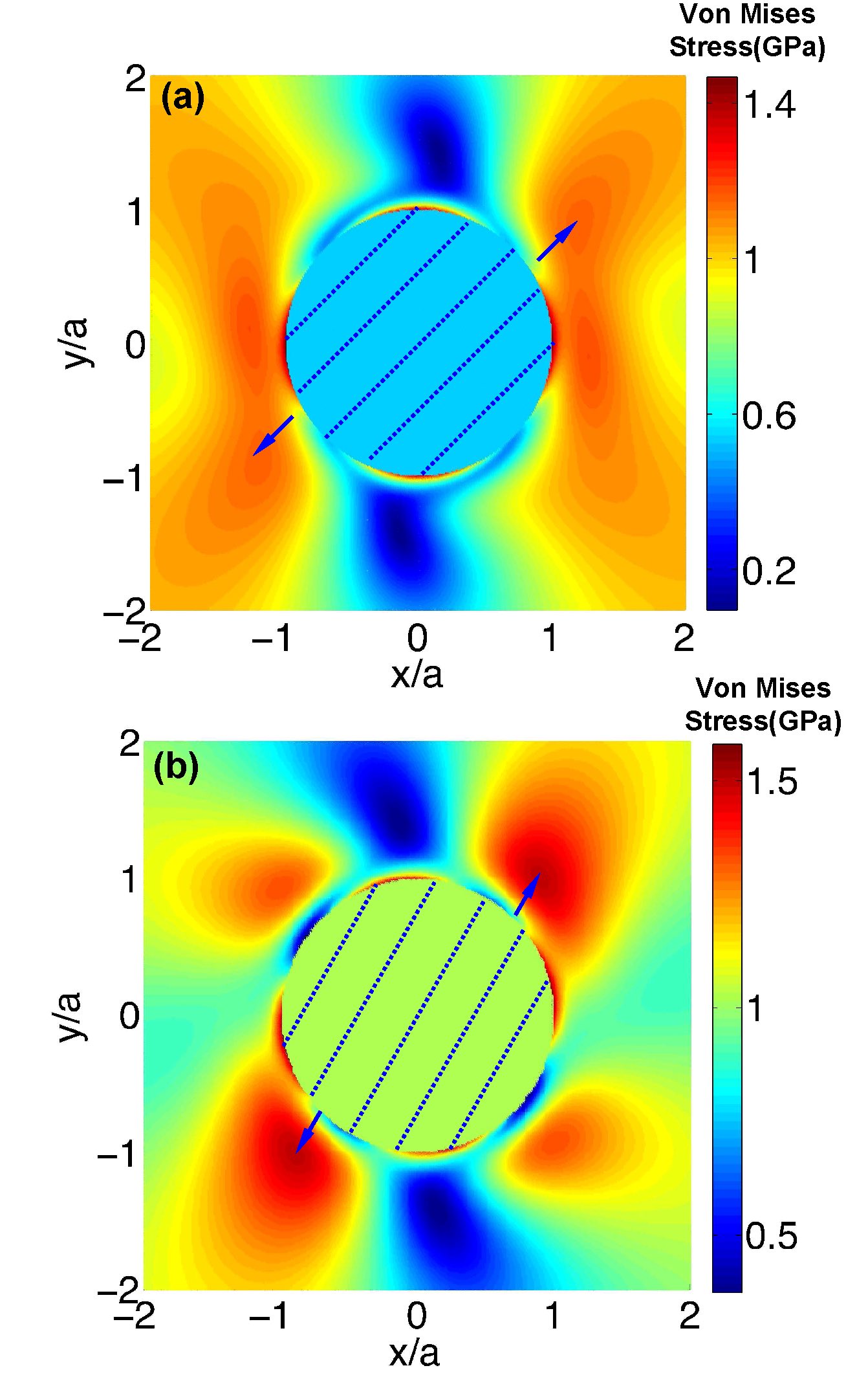}
\caption{\label{fig8} The von Mises stress ($\sigma^{v}$) distribution due to the MT in the vicinity region of spherical phase of $x-y$ plane. Two mappings with $\theta=45^{\circ}$ (a) and$\theta=60^{\circ}$(b) are showed here to track the variation of $\sigma^{v}$ with the MT shear direction. The arrows indicated the regions with maximum von Mises stress.}
\end{figure}

\textcolor{blue}{Figure \ref{fig8}}(a)-(b) display the contour plots of the von Mises stress $\sigma^{v}$ in the vicinity of a transforming spherical B2 phase for two shear angles (a) $\theta=45^{\circ}$ and (b) $\theta=60^{\circ}$. From these figures, one can see clearly that $\sigma^{v}$ is constant inside the crystalline phase but exhibits strong position dependence in the glassy matrix, as consistent with the Eshelby theory\citep{Eshelby}. As can be seen, $\sigma^{v}$ is small (generally less than 1 GPa) in the regions straying away from the direction of the martensitic transformation, such as those colored in blue, while becomes intensified in the regions aligning with the direction of the martensitic transformation, such as those colored in red. This is particularly so for the high transformation angle $\theta=60^{\circ}$, at which the maximum value of $\sigma^{v}$ could exceed the macroscopic yield stress of the glassy matrix ($\sim$1.5 GPa) in the directions close to that of phase transformation (as indicated by the arrows in \textcolor{blue}{Fig. \ref{fig8}}(b)). This indicates that these regions are the preferable sites for shear-band initiation in the glassy matrix during the martensitic transformation. Since the maximum $\sigma^{v}$ always appear in the direction aligning with the transformation angle $\theta$, the formation of the type II shear bands, as observed in our experiments (\textcolor{blue}{Fig. \ref{fig5}}), can be rationalized as a direct result of the stress redistribution in the glassy matrix during the martensitic transformation of the B2 phases.

Next, we would like to calculate the stress/strain distribution in the BMG composite after the martensitic transformation of the B2 phases. For a uniaxial compressive stress ($\sigma^{A}<0$) along the y axis, the effective ''transformation'' strain or eigenstrain $\epsilon^{T}_{ij}$ can be written as: 
             \begin{equation}\label{eq5} 
             \epsilon_{ij}^{T}=(A\sigma_{A}/9K)\delta_{ij}+(B\sigma_{A}/6G)(3n_{i}n_{j}-\delta_{ij})  
             \end{equation}   
where $A$ and $B$ are the two coefficients associated with the elastic mismatch between the martensitic phase and the glassy matrix. Note that the uniaxial loading along the $y$ axis corresponds to $n_{1}=n_{3}=0$ and $n_{2}=1$; the first term in the right hand side of \textcolor{blue}{Eq.}\ref{eq5} is for pure dilation while the second term for pure shear. Likewise, for the pure shear term in \textcolor{blue}{Eq.}\ref{eq5}, the resultant elastic stress/strain field can be derived following a very similar procedure as that used for the transience of the martensitic transformation of the B2 phases; while for the pure dilation term $\epsilon_{kl}^{T}=\epsilon_{d}\delta_{kl}$, where $\epsilon_{d}$ is the magnitude of the dilatational strain, an explicit expression for the corresponding stress/strain field could be also obtained (see \textcolor{blue}{Eq.}\ref{C1}).  Afterwards, based on the principle of superposition in linear elasticity, one can obtain the elastic strain/stress field and thus the von Mises stress distribution in the presence of a martensitic phase in the glassy matrix. At the present time, there are no available data reported for the elastic modulus of the CuZr martensitic phase. However, according to the measured hardness data\citep{ADMA:ADMA201000482}, the elastic modulus of the martensitic phases  should be larger than that of the B2 phases. As a result, this yields the negative values of $A$ and $B$, which means that a compressive external strain $\epsilon^{A}<0$ always corresponds to a tensile effective eigenstrain ($\epsilon^{T}>0$), and \textit{vice versa}. As will be shown in the later text, this behavior has an important effect on the stability of the particle-glass interface and thus the ductility of the CuZr BMG composite under different loading condition (compression or tension). 
                                                                                                                                                             
\begin{figure}\centering
\includegraphics[width=100mm]{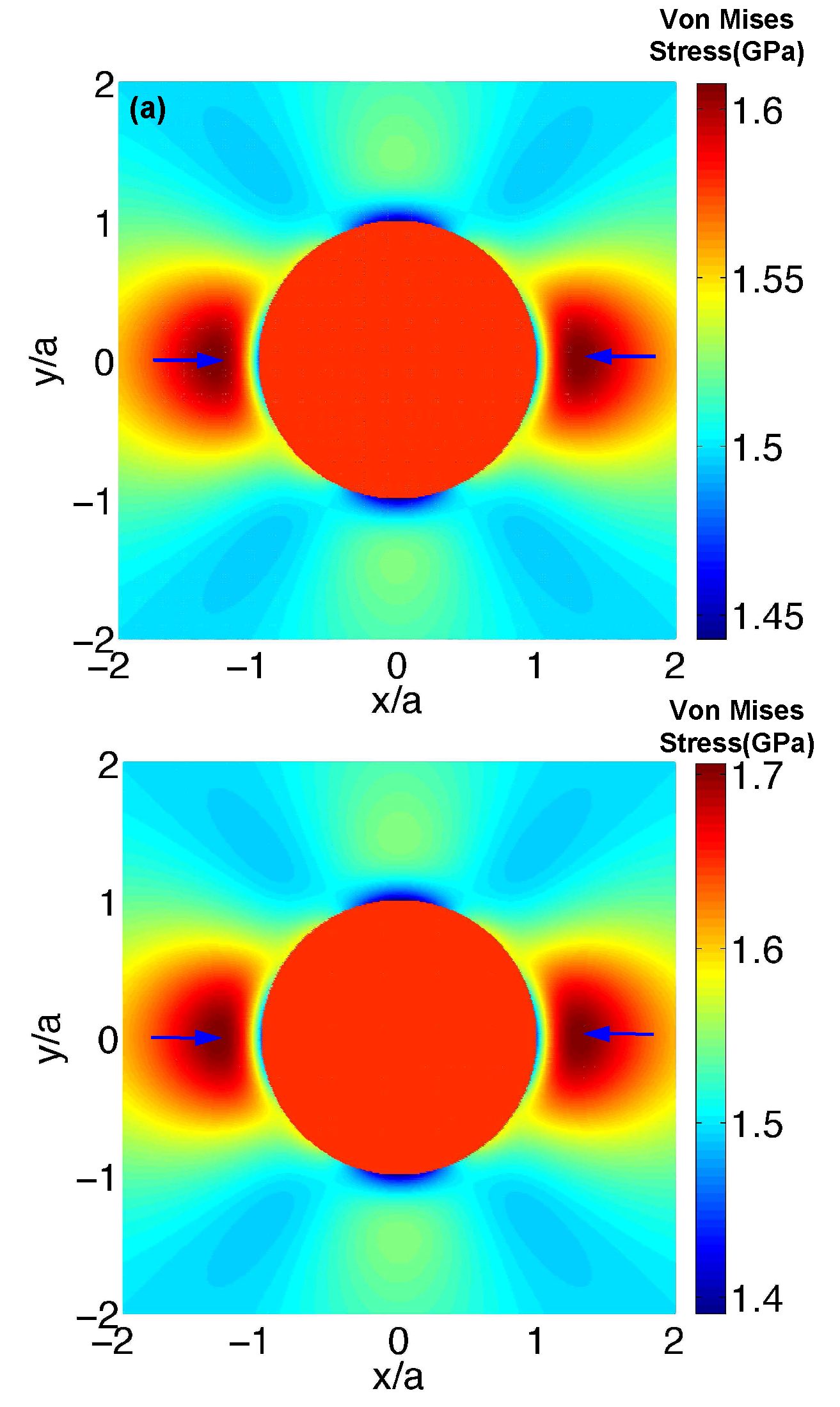}
\caption{\label{fig9} The von Mises stress ($\sigma^{v}$) distribution due to the constraint of glassy matrix on deformation of martensite phases  in the vicinity region of a spherical phase of $x-y$ plane. Two mappings with $x=0.1$(a) and $x=0.2$(b) are showed here.  The arrows indicated the regions with maximum von Mises stress.}
\end{figure}

    According to \textcolor{blue}{Eq.}\ref{eq4} the values of $A$ and $B$ are a function of the elastic constant difference $x$ [$x=(C_{1}-C)/C$, $C$ denotes $K$ or $G$]. Here, we take the typical values of $A$ and $B$ at $x=0.25$ and 0.50 and calculate the corresponding von Mises stress distribution in the $x-y$ plane, as shown in \textcolor{blue}{Fig. \ref{fig9}} (a) and (b) respectively. In this case, the maximum values of $\sigma^{v}$ appear close to the martensitic phase and the direction of the maximum von mises stress is perpendicular to the loading axis, as indicated by the arrows in \textcolor{blue}{Fig. \ref{fig9}}(a) and (b). In addition, the variation of $x$ only changes the intensity of $\sigma^{v}$, but the distribution of $\sigma^{v}$ remains the same. This behavior indicates that, as the external load continues increasing after the martensitic transformation, the elastic mismatch between the martensitic phase and the glassy matrix will cause shear-band initiation in the direction vertical to the loading axis, which is exactly what we observed in the experiments, that is, the type III shear bands as seen in \textcolor{blue}{Fig. \ref{fig5}}. 
 
\subsection{Mechanistic mechanism for the strain hardening ability}
 Compared to monolithic BMGs, one of prominent deformation features of Cu-Zr based BMG composites is their strain or work hardening ability\citep{Hofmann10092010,ADMA:ADMA201000482}. According to the prior works\citep{Pauly20104883,Hofmann10092010,ADMA:ADMA201000482}, the observed work hardening in the CuZr based BMG composites cannot be solely attributed to the intrinsic strain hardening of the B2 phases, but also arises from a constraining effect of the glassy matrix on the martensitic transformation and the subsequent deformation of the transformed phases. In theory, this constraint effect, also called Eshelby back stress effect, increases the elastic energy stored in the whole composite system, which leads to an increase in the applied stress and thus manifests as strain hardening in the BMG composite. Now, with the obtained expressions for the various stress/strain fields, the contribution to the strain hardening from this constraint effect can be quantitatively analyzed. According to Eshelby's theory\citep{Eshelby}, the total increase in the elastic energy storage $E_{1}$ in the BMG composite during the martensitic transformation, i.e. for a spherical inclusion undergoing a pure shear $\epsilon^{T}_{ij}=\epsilon^{T}(2n_{i}n_{j}-1)$($i, j=1,2$ and $n_{3}=0$), can be written as: 

             \begin{equation}\label{eq6} 
             E_{1}=\frac{2(7-5\nu)}{15(1-\nu)}G(\epsilon^{T})^{2}V+\sigma^{A}\epsilon^{T}\phi V
                          \end{equation}   
where $V$ is the volume of the inclusion. Note that the first term at the right-hand side of \textcolor{blue}{Eq.}\ref{eq6} is the strain energy stored due to the martensitic transformation of the inclusion and the second term is the strain energy stored due to the interaction of the elastic field of the transforming inclusion and the external stress field, i.e. the uniaxial far-field stress $\sigma^{A}$ along the $y$ axis, which can be generally expressed as $\int_{V}\sigma_{ij}^{A}\epsilon_{ij}^{T}dv$\citep{Eshelby}. Since the only non-zero component of $\sigma_{ij}^{A}$ is $\sigma_{22}^{A} $($=\sigma_{A}$), $\int_{V}\sigma_{ij}^{A}\epsilon_{ij}^{T}dv=\sigma^{A}\epsilon^{T}_{22}V$ with $\epsilon^{T}_{22}=\epsilon^{T}\phi(\mathbf{n}, \mathbf{m},\mathbf{k})$, where $\phi$ is a parameter that is a function of the orientation of the B2 phase. It should be noted here that the elastic energy $E_{1}$ is instantaneous and vanishes after the phase transformation. On the other hand, after the phase transformation, the strain energy stored $E_{2}$ due to the interaction of the elastic field of the inhomogeneity with the external uniaxial stress field is\citep{WeibergerC}: 
                       \begin{equation}\label{eq7} 
                E_{2}=-\frac{1}{2}\int_{V}\sigma_{ij}^{A}\epsilon_{ij}^{T}dv=-\frac{1}{2}(\sigma^{A})^{2}(\frac{A}{9K}+\frac{B}{3G})
                                        \end{equation}                                
Note that the negative sign in the above expression makes $E_{2}$ positive because the coefficient $A$ and $B$ are both negative.
                                        
\begin{figure}\centering
\includegraphics[width=120mm]{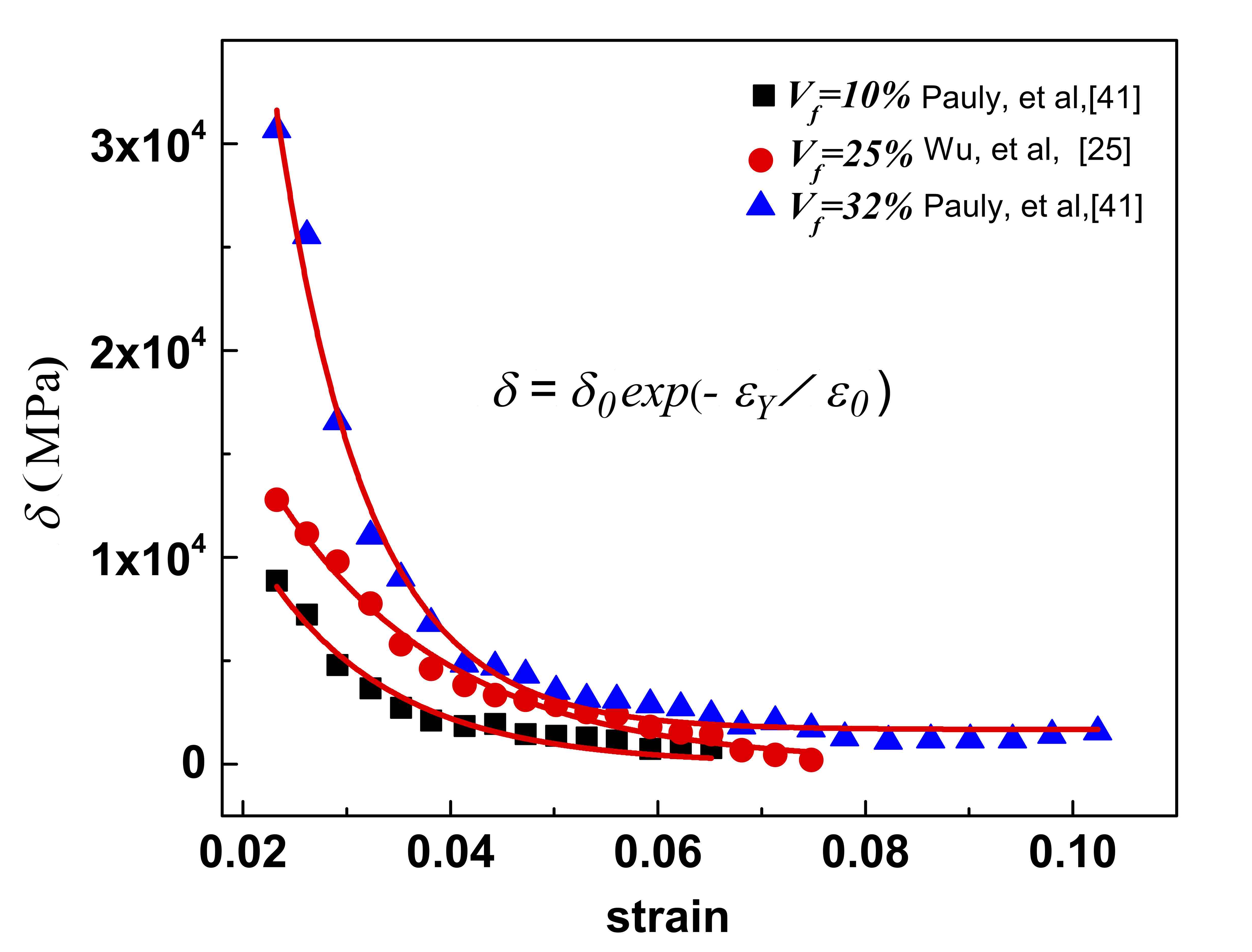}
\caption{\label{fig10} The variation of the work-hardening coefficient  with the applied strain for different volume fraction of B2 phases collected from literatures. These data points are well fitted by the exponential decaying function $\delta=\delta_{0}exp(-\epsilon_{y}/\epsilon_{0})$. The values of fitting parameters $\delta_{0}=5.6\times10^{4}$ MPa, $\epsilon_{0}=0.01$ for $V_{f}=10$\%; $\delta_{0}=5.3\times10^{4}$ MPa, $\epsilon_{0}=0.017$ for $V_{f}=25$\% and $\delta_{0}=4.2\times10^{5}$ MPa, $\epsilon_{0}=0.008$ for $V_{f}=32$\%}
\end{figure}

    Assume that the number of the transformed B2 phases is $Nf$, where $N$ is the total number of the B2 phase per volume in the undeformed sample and $f$ is the accumulated percentage of transformation. Hence, the number of the B2 phases undergoing the martensitic transformation can be written as $NP(1-f)$, where $P$ is the instantaneous probability for a B2 phase to undergo the martensitic transformation. As a result, the total elastic energy stored is:
                                        \begin{equation}\label{eq8} 
                E_{total}=NP(1-f)\bar{E_{1}}+NfE_{2}
                                        \end{equation}       
 where $\bar{E_{1}}$  is the elastic energy $E_{1}$ averaged over all possible orientations of the B2 phases. The accumulated fraction of the transformed B2 phases, $f=f(\epsilon_{y})$, is assumed to be a single-valued function of the external strain $\epsilon_{y}$. As a result of the additional elastic energy storage (\textcolor{blue}{Eq.}\ref{eq8}), the applied stress has to increase, resulting in strain hardening.  The strain hardening coefficient $\delta$ is defined as $\delta=d\sigma/d\epsilon_{y}=d^{2}E_{total}/d\epsilon_{y}^{2}$. From \textcolor{blue}{Eq.}\ref{eq8}, we could obtain $\delta=N(E_{2}-P\bar{E_{1}})(d^{2}f/d\epsilon_{y}^{2})$. From the simple linear-growth model, one infer that the transformation rate $df/d\epsilon_{y}=(1-f)/\epsilon_{0}$, i.e., the transformation rate of the B2 phases be proportional to the volume fraction $(1-f)$ of un-transformed B2 phases and $\epsilon_{0}$ denotes a reference strain. Solving the above equation gives $f(\epsilon_{y})=1-exp(-\epsilon_{y}/\epsilon_{0})$.  Combining this equation with the expression for $\delta$, we finally obtain $\delta=\delta_{0}exp(-\epsilon_{y}/\epsilon_{0})$with $\delta_{0}=N(P\bar{E_{1}}-E_{2})/\epsilon_{0}^{2}$.  To verify our modeling, we collected the strain hardening coefficients as a function of the plastic strain from the stress-strain curves, as reported in the literature\citep{ADMA:ADMA201000482,SPaulyapl}, for the CuZr-based BMG composites with different volume fractions of B2 phases. As show in \textcolor{blue}{Fig. \ref{fig10}}, it is evident that our theoretical modeling is in excellent agreement with the experimental data, which indicates that the strain hardening ability of the Cu-Zr based metallic glass composites can be attributed to the Eshelby back stress effect, as proposed in the previous works\citep{Pauly20104883}. Since the volume fraction of the B2 phases is proportional to $N$, we can thus infer, the strain hardening coefficient should increase with the volume fraction of the B2 phases in a linear way, which is consistent with the previous studies\citep{songaioadv}. In addition, the dependence of the strain hardening coefficient $\delta_{0}$ on the term $\delta_{0}=N(P\bar{E_{1}}-E_{2})/\epsilon_{0}^{2}$ implies that one has to ensure $NP\bar{E_{1}}>E_{2}$ in order to retain a positive strain hardening coefficient. In other words, in terms of achieving the strain hardening ability and tensile ductility in the BMG composites, it is beneficial to reduce the extent of the elastic mismatch (as for a low $E_{2}$) and/or to design the orientation of the B2 phases for a high and efficient instantaneous transformation probability $P$.
 
\subsection{Stress-affected zones (SAZ) of crystalline phases and their effects on ductility}
 
\begin{figure}\centering
\includegraphics[width=120mm]{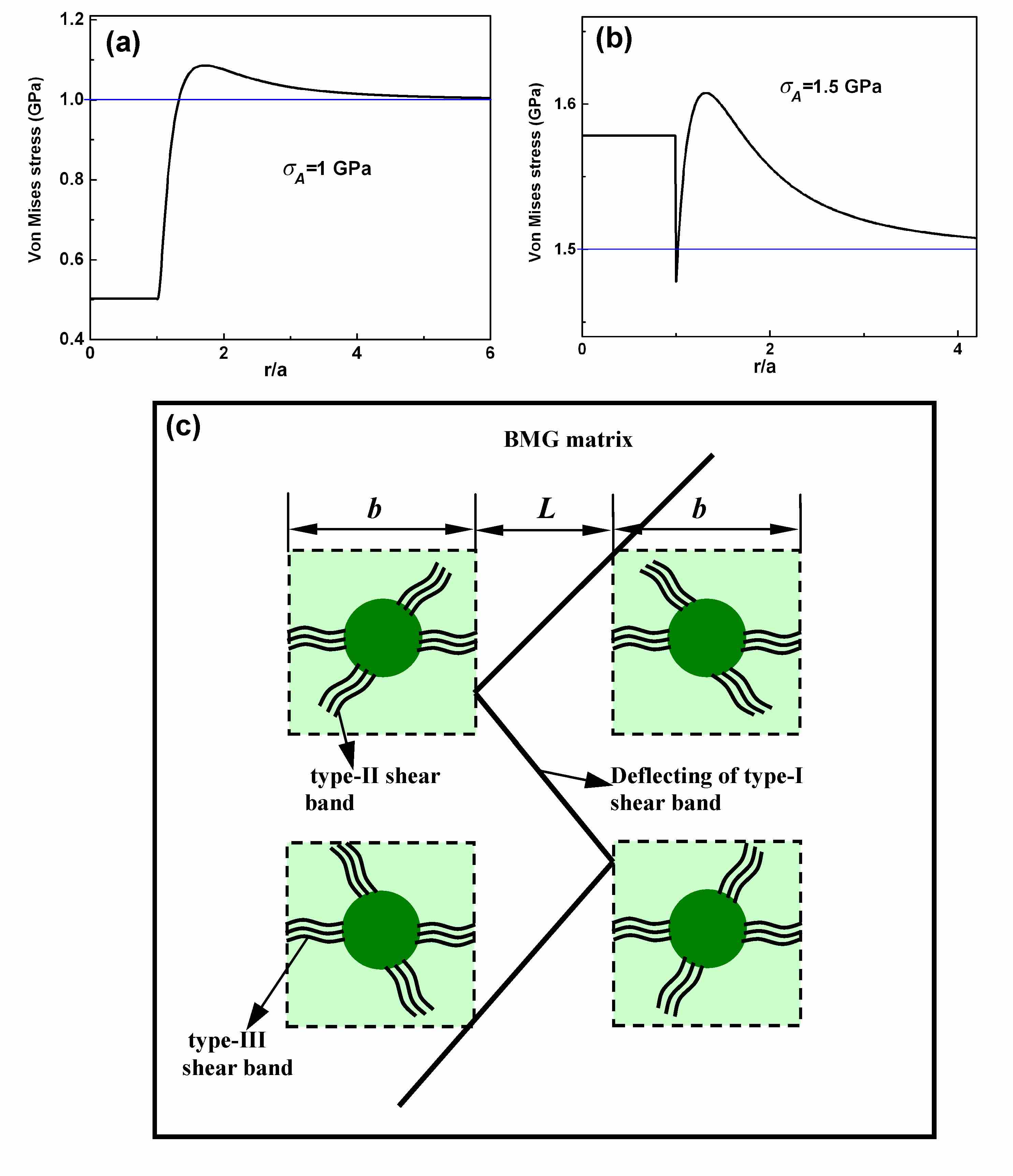}
\caption{\label{fig11} Schematic diagram illustrating the stress affected zone (SAZ) due to the stress redistribution of the MT of B2 phases and the subsequent deformation of martensite phases and their deflection on the type I shear band during the deformation of CuZr based metallic glass composites.}
\end{figure}

 Based on the experimental results (\textcolor{blue}{Fig. \ref{fig5}}) and theoretical analyses, it is clear that there is a stress-affected zone (SAZ) for every B2 phase embedded in the glassy matrix, within which the martensitic transformation can lead to the formation of type II and III shear bands. From an experimental viewpoint, the extension of these shear bands could be used to estimate the SAZ size. Theoretically, we can assume the SAZ to be a cubic region for simplicity and the side length b of the SAZ can be estimated from the decay of the constraining stress with the distance $r$ from the center of the B2 phase.  \textcolor{blue}{Figure \ref{fig11}} (a) and (b) show the variation of the von-Mises stress  with $r$ along the propagation direction of type II ($\theta=60^{\circ}$) and III shear bands, respectively. As seen in these figures, the von Mises stress varies significantly in the proximity of the crystalline phase but approaches to the far field stress far away. For type II shear bands, $\sigma^{v}$ reaches the external stress level when the distance $r> 3.5a$, which corresponds to $b\sim2.5a$; while for type III shear bands, this distance can be set as $2.5a$, which also corresponds to $b\sim2.5 a$. Therefore, we can safely choose $b=2.5a$ as a theoretical estimate of the SAZ size, beyond which the back stress effect caused by the martensitic transformation can be neglected. 

 The distribution of the SAZs in the glassy matrix could have a significant effect on the deformation stability and ductility of the Cu-Zr based metallic-glass composites. For a composite with a dilute B2 concentration, the final fracture of the composite can take place along the path of the type I shear band as caused by the external stress. Therefore, the stability of the type I shear bands determines the overall ductility of the metallic-glass composite, as in monolithic metallic glasses\citep{Song2008813,Han20091367,DallaTorre20103742}. However, a large population of SAZs could deflect the propagation of the type I shear band, thus delaying the occurrence of shear-banding instability as illustrated in \textcolor{blue}{Fig. \ref{fig3}}(b) and \textcolor{blue}{Fig. \ref{fig11}}(c). Since the type I shear band propagates through the regions between the SAZs, the smaller is the spacing $L$ between two adjacent SAZs the more effectively is SAZs in shear-band deflection. For a homogeneous distribution of the crystalline phases, $L$ can be estimated as $b=1/\sqrt[3]{V_{f}/(4\pi a^{3}/3)}-b$ by taking the composite sample to be a cubic shape with a unit length, where $V_{f}$ is the volume fraction of the crystalline phases. From this expression, one can see that $L$ decreases with the increasing $V_{f}$, which indicates that a higher volume fraction of crystalline phases can lead to a better ductility of the metallic-glass composite, as consistent with the reported experimental results\citep{SPaulyapl}. However, as $V_{f}$ increases and the SAZs begin to overlap, $L$ will finally become zero when $V_{f}$ reaches a critical value. When this happens, the deflection of the type-I shear bands is no longer the controlling mechanism for the deformation stability of the composite. Other mechanisms, such as interface debonding, may come into effect. Taking $b=2.5a$ into the expression for $L$, one obtains the critical volume fraction $V_{fc}\approx27\%$. Interestingly, this value of $V_{fc}$ is very close to the percolation threshold ($\sim$30\%) as experimentally identified for the crystalline phases in the Cu-Zr metallic-glass composites\citep{SPaulyapl}. 
 
 On the other hand, for a give $V_{f}$, $L$ decreases with the decreasing radius of the B2 phases. This implies a size effect of the crystalline phase on the ductility of the metallic-glass composites. Given a fixed volume fraction, the smaller is the size of the individual crystalline phase the denser is its spatial distribution. In turn, this leads to more effective deflection of the type-I shear bands and thus better ductility. However, to enhance the ductility of the metallic-glass composite, it can be envisioned that the size of the crystalline phases cannot be reduced infinitely because of the finite thickness of a shear band, which ranges from 10 nm to 100 nm[5]. When the size of a crystalline phase is comparable to the thickness of a shear band, it becomes less effective to deflect the propagation of the type-I shear band. Consequently, one may foresee an optimum size for the crystalline phase to enhance the overall ductility of the metallic-glass composites, the details of which deserve further theoretical and experimental studies. 
 
\subsection{Stability of amorphous/crystalline interface}
 
  Finally, we would like to consider the issue of the stability of the interface between the crystalline phase and the glassy matrix. The debonding of the interface has been observed during the final deformation stage of a Cu-Zr metallic glass composite under tensile loading\citep{ADMA:ADMA201000482}. In principle, the crystalline-amorphous interface is considered stable under a compressive normal stress or a negative hydrostatic pressure ($\sigma_{ii}<0$) but may debond under a tensile normal stress or a positive hydrostatic pressure ($\sigma_{ii}>0$).  Now let us calculate the hydrostatic pressure on the crystalline-amorphous interface caused by the martensitic transformation. According to the Eshelby theory\citep{Eshelby}, the mechanical strain across the inclusion-matrix interface satisfies the following relation: 
                        \begin{equation}\label{eq9} 
  \epsilon_{ii}^{C}(out)=\epsilon_{ii}^{C}(in)-\frac{1}{3}\frac{1+\nu}{1-\nu}\epsilon_{ii}^{T}-\frac{1-2\nu}{1-\nu}('\epsilon^{T}_{ij}g_{i}g_{j})
                                        \end{equation}                                                       
where $\epsilon_{ii}^{C}(out)$ and $\epsilon_{ii}^{C}(in)$ denote the values of $\epsilon_{ii}^{C}$ in the matrix and inclusion, respectively, as across the interface; $g_{i}$ is the component of the unit vector $\mathbf{g}=\mathbf{r}/r$ with $\mathbf{r}$ being the vector ending at the matrix-inclusion interface. For the martensitic transformation of the B2 phase,$\epsilon_{ii}^{T}=0$, $\epsilon_{ii}^{C}(in)=S_{iikl}\epsilon_{kl}^{T}=0$, $'\epsilon_{ij}^{T}=\epsilon_{ij}^{T}$ and thus $\epsilon_{ii}^{C}(out)=-\frac{1-2\nu}{1-\nu}\epsilon^{T}_{ij}g_{i}g_{j}$. Therefore, the interfacial hydrostatic pressure can be expressed as $p=K\epsilon_{ii}^{C}(out)=-\frac{1-2\nu}{1-\nu}K\epsilon^{T}_{ij}g_{i}g_{j}$, which varies along the periphery of the crystalline-amorphous interface. Based on the analysis in the \ref{appendixc}, the maximum tensile pressure ($p^{T}_{max}$) is attained at the interfacial point with the vector $\mathbf{r}$ coincident with the principal axis $\mathbf{m}$ of $\epsilon_{ij}^{T}$, which reads $p_{T}^{max}=\frac{1-2\nu}{1-\nu}K\epsilon^{T}(1+\xi)$, where $\xi=\lambda_{n}/(2\lambda_{n}+\lambda_{k})$ for $\lambda_{n}>\lambda_{k}>\lambda_{m}$. While the maximum compressive pressure ($p_{max}^{C}$) is attained at the interfacial point with the vector $\mathbf{r}$ perpendicular to the principal axis $\mathbf{m}$ of $\epsilon_{ij}^{T}$, which reads $p_{C}^{max}=-\frac{1-2\nu}{1-\nu}K\epsilon^{T}(1-\psi)$ where $\psi=\lambda_{k}/(2\lambda_{n}+\lambda_{k})$.  With these equations, we can roughly estimate the interfacial bonding strength between the crystalline phases and the amorphous matrix. This can be done by equating the maximum tensile hydrostatic pressure $p_{max}^{T}$ to the interfacial bonding strength $\sigma_{inter}$. By taking $\epsilon^{T}\sim0.01$, $\xi\sim0.5$, $K=114$ GPa and $\nu=0.37$, the estimated interfacial bonding strength $\sigma_{inter}=700$ MPa. This is slightly lower than the yield strength of the metallic-glass composite. In principle, based on the theoretical framework, we can extract the interfacial bonding strength if the value of $\epsilon^{T}$ is exactly known or vice versa.

Following a similar procedure, the hydrostatic pressure acting on the crystalline-amorphous interface after the martensitic transformation can be derived as $p=-\frac{1-2\nu}{1-\nu}(KB\sigma^{A}/6G)(3n_{i}n_{j}-\delta_{ij})g_{i}g_{j}$, where the principal axis of the effective transformation strain coincides with the external $x-y-z$ axis, such that $n_{1}=n_{3}=0$, $n_{2}=1$.  For the case of uniaxial compression ($\sigma^{A}<0$), the the maximum tensile pressure $p_{T}^{max}=\frac{1-2\nu}{1-\nu}KB\sigma^{A}/6G$ is attained at the points where the interface intersects the $x$-axis or $z$-axis; while the maximum compressive pressure $p_{C}^{max}=\frac{1-2\nu}{1-\nu}KB\sigma^{A}/3G$ is attained at the points where the interface intersects the $y$-axis. For the case of uniaxial tension, the sign of the above pressures is changed but the magnitude remains the same. Taking $B\sim0.1$ $\sigma^{A}=1.5$ GPa, $K=114$ GPa, $G=32$ GPa, $\nu=0.37$, the estimated maximum tensile pressure is 37 MPa, which is much smaller than that due to martensitic transformation of B2 phases.  Thus, the hydrostatic pressure due to the elastic mismatch between the martensite phase and the glassy matrix can be reasonably neglected in the estimation of the interface bonding strength.

 \section{Summaries and conclusions}
 In summary, salient conclusions can be drawn below based on our current experimental and theoretical studies:

\begin{enumerate}[(1)]
   \setlength{\itemindent}{2.0em}
    \item Three types of shear bands with distinct morphological characteristics are identified in the deformed metallic-glass composites. The type-I shear bands are oriented 45 degrees relative to the loading axis and related to the external stress field; while the type-II and type-III shear bands as seemingly joining or emanating from the crystalline phases have a different orientation. Based on the Eshelby theory, it is shown that the type-II shear bands result from the stress field caused by the martensitic transformation of the B2 phases, while the type-III shear bands result from the stress field caused by the elastic mismatch induced after the martensitic transformation.
    \item The back stress effect during the martensitic transformation and the elastic misfit effect afterwards act together and increase the elastic energy storage during the deformation of the metallic-glass composite, giving rise to work hardening.
    \item A stress-affected zone (SAZ) around the crystalline phase is defined based on our stress analysis. The presence of the SAZs rationalizes the phenomenon of the deflection of the type-I shear bands.   
    \item The hydrostatic pressure acting on the crystalline-amorphous interface is derived. Depending on the loading modes, the maximum tensile pressure could be attained at different points along the interface. Interfacial debonding may take place once the tensile hydrostatic pressure exceeds the interfacial bonding strength. 
   
   \end{enumerate}
 
 \section*{Acknowledgements}
 Experimental assistance and insightful discussion with  Prof Z. P. Lu, Dr. Q.Wang, , Dr. Z. Y. Liu, Dr. W. Jiao and Mr. Y. F. Ye, Mrs Y.M. Lu are appreciated. The work is supported by German Science Fundation under Leibniz Program (Grant No. EC111/26-1) and the Research Grant Council (RGC) of the Hong Kong government through the General Research Fund (GRF) with the account numbers CityU117612, 102013, 11209314, 9042066 and 9054013.
 
 \appendix
 \section{Derivation of the displacement field and the strain field for a spherical Eshelby inclusion}\label{appendixa}
 According to the Green's function theory, the displacement at the position $\mathbf{r}$ due to a point force $F_{i}$ at $\mathbf{r'}$ in an infinite elastic medium is:
  \begin{equation}\label{A1} 
  \setlength{\abovedisplayskip}{24pt}
\setlength{\belowdisplayskip}{24pt}
 u_{j}(\mathbf{r}-\mathbf{r'})=\frac{1}{4\pi G}\frac{F_{j}}{|\mathbf{r}-\mathbf{r'}|}-\frac{1}{16\pi G(1-\nu)}F_{l}\frac{\partial^{2}}{\partial x_{l}x_{j}}|\mathbf{r}-\mathbf{r'}| 
   \end{equation} 
where $G$ and $\nu$ are the shear modulus and Poisson's ratio, respectively. According to \textcolor{blue}{Eq.}\ref{A1} and superposition principle of linear elasticity, the constrained displacement in the stage III as described in section 3 is:
\begin{equation}\label{A2} 
 u_{i}^{C}=\int_{S}dS_{k}\sigma^{T}_{ik}U_{i}(\mathbf{r}-\mathbf{r'})
    \end{equation}  
where the integration is performed on the surface of the spherical B2 phase $\mathbf{S}$. Using the Gauss theorem and $\partial x_{i}=-\partial x_{i}'$  when acting on the $|\mathbf{r}-\mathbf{r'}|$, one obtains:
          \begin{equation}\label{A3} 
 u_{i}^{C}=\int_{v}dv\sigma^{T}_{ik}U_{i,k}(\mathbf{r}-\mathbf{r'})=-\frac{1}{4\pi G}\sigma^{T}_{ik}\phi_{,k}+\frac{1}{16\pi G(1-\nu)}\sigma_{jk}^{T}\psi_{,ijk}
    \end{equation}
where $\psi=\int_{v}|\mathbf{r}-\mathbf{r'}|dv$ and $\phi=\int_{v}\frac{dv}{|\mathbf{r}-\mathbf{r'}|}$ are the biharmonic potential and the Newtonian potential of attracting matter of unit density filling the inclusion volume $v$.

For a spherical B2 phase, we use the spherical coordinate to integrate $\psi$ and $\phi$, respectively. In the integration, we choose the $z$ axis always aligning along the direction of vector $\mathbf{r}$. By doing this, the $\psi$ and $\phi$ is integrated as:

\begin{subequations}\label{A4}
\begin{eqnarray}
\psi&=&\int_{v}|\mathbf{r}-\mathbf{r'}|dv=\int_{0}^{a}\int_{0}^{\pi}\int_{0}^{2\pi}\sqrt{r^{2}+r'^{2}-2rr'\cos{\theta}}r'^{2}\sin{\theta}dr'd\theta d\varphi \nonumber \\  \nonumber\\
&=&2\pi\int_{0}^{a}\left[2rr'^{2}+(2/3)(r'^{4}/r)\right]dr'=\frac{4\pi r}{3}a^{3}+\frac{4\pi}{15r}a^{5} \\  \nonumber\\
\phi&=&\int_{v}\frac{dv}{|\mathbf{r}-\mathbf{r'}|}=\int_{0}^{a}\int_{0}^{\pi}\int_{0}^{2\pi}\frac{r'^{2}\sin{\theta}dr'd\theta d\varphi}{\sqrt{r^{2}+r'^{2}-2rr'\cos{\theta}}}\nonumber \\ \nonumber\\
&=&2\pi\int_{0}^{a}\int_{0}^{\pi}\frac{-r'd(r'\cos{\theta})d\theta}{\sqrt{r^{2}+r'^{2}-2rr'\cos{\theta}}}=\frac{4\pi a^{3}}{3r}
\end{eqnarray}
\end{subequations} \\                            
      
Based on \textcolor{blue}{Eqs.}\ref{A4}, we have:

\begin{subequations}\label{A5}
\begin{eqnarray}
\psi_{,i}&=&\frac{4\pi a^{3}}{3}\frac{x_{i}}{r}-\frac{4\pi a^{5}}{15}\frac{x_{i}}{r^{3}}\\ \nonumber\\
\psi_{,ij}&=&\frac{4\pi a^{3}}{3}\left(\frac{\delta_{ij}}{r}-\frac{x_{i}x_{j}}{r^{3}}\right)-\frac{4\pi a^{5}}{15}\left(\frac{\delta_{ij}}{r^{3}}-\frac{3x_{i}x_{j}}{r^{5}}\right)\\ \nonumber\\
\psi_{,ijk}&=&\frac{4\pi a^{3}}{3}\left(-\frac{x_{i}\delta_{jk}+x_{i}\delta_{ik}+x_{k}\delta_{ij}}{r^{3}}+\frac{3x_{i}x_{j}x_{k}}{r^{5}}\right)+\frac{4\pi a^{5}}{15}\left(\frac{3(x_{i}\delta_{jk}+x_{i}\delta_{ik}+x_{k}\delta_{ij})}{r^{5}}-\frac{15x_{i}x_{j}x_{k}}{r^{7}}\right)\nonumber\\ \\
\phi_{,k}&=&-\frac{4\pi a^{3}}{3}\frac{x_{k}}{r^{3}}
\end{eqnarray}
\end{subequations}      
                                                                                  
Taking \textcolor{blue}{Eqs.}\ref{A5} into \textcolor{blue}{Eq.}\ref{A3}, we get the constrained displacement due to the spherical inclusion:
                                       
\begin{equation}\label{A6} 
 u_{i}^{C}=\frac{1}{6G(1-\nu)}\left[(1-2\nu)\frac{a^{3}}{r^{3}}+\frac{3}{5}\frac{a^{5}}{r^{5}}\right]\sigma^{T}_{ij}x_{j}+\frac{1}{4G(1-\nu)}\frac{a^{3}}{r^{5}}\left(1-\frac{a^{2}}{r^{2}}\right)\sigma_{jk}^{T}x_{i}x_{j}x_{k}
 \end{equation}  \\     
and the constrained strain $\epsilon_{ij}^{C}$ (as shown in \textcolor{blue}{Eq.}\ref{eq2} in the text) is readily obtained from $\epsilon_{ij}^{C}=(u_{i,j}^{C}+u_{j,i}^{C})/2$.

\section{Explicit expressions for the constrained strain field due to the MT and the presence of martensite phase in the matrix}\label{appendixb}
 
   For ($i,j=1,2$), there are following equivalences\citep{ProcacciaPRL}:
 \begin{subequations}\label{B1}
\begin{eqnarray}
\epsilon_{ij}^{T}x_{j}&=&\epsilon^{T}[2n_{i}(\mathbf{n}\centerdot\mathbf{r})-x_{i}]\\ 
x_{i}\epsilon_{ij}^{T}x_{j}&=&x_{i}\epsilon^{T}(2n_{i}n_{j}-\delta_{ij})x_{j}=\epsilon^{T}[(\mathbf{n}\centerdot\mathbf{r})^{2}-r^{2}]
\end{eqnarray}
\end{subequations}         
   
   Taking these equivalences into \textcolor{blue}{Eq.}\ref{eq2} and noticing that $\sigma_{ij}^{T}=2G\epsilon_{ij}^{T}$, one could get the explicit expression for the constrained strain field due to a pure shear of eigenstrain in $x-y$ plane:
    
\begin{equation}\label{B2}
\begin{aligned}
&\epsilon_{ij}^{C}=\frac{\epsilon^{T}}{5(1-\nu)}\left(\frac{a^{5}}{r^{5}}\right)\left\{\frac{5x_{i}x_{j}[2(\mathbf{n}\centerdot\mathbf{r})^2-r^{2}]-2r^{2}[(\mathbf{n}\centerdot\mathbf{r})(n_{i}x_{j}+n_{j}x_{i})-x_{i}x_{j}]}{r^{4}}\right\}+\\ \\
&\frac{\epsilon^{T}}{3(1-\nu)}\left(\frac{a^{3}}{r^{3}}\right)\left[(1-2\nu)+\frac{3}{5}\left(\frac{a^{2}}{r^{2}}\right)\right]\left\{\frac{2r^{2}(2n_{i}n_{j}-\delta_{ij})-6[(\mathbf{n}\centerdot\mathbf{r})(n_{i}x_{j}+n_{j}x_{i})-x_{i}x_{j}]}{2r^{2}}\right\}+\\ \\
&\frac{\epsilon^{T}}{2(1-\nu)}\left(1-\frac{a^{2}}{r^{2}}\right)\left(\frac{a^{3}}{r^{3}}\right)\left\{\frac{(r^{2}\delta_{ij}-10x_{i}x_{j})[2(\mathbf{n}\centerdot\mathbf{r})^{2}-r^{2}]+2r^{2}[(\mathbf{n}\centerdot\mathbf{r})(n_{i}x_{j}+n_{j}x_{i})-x_{i}x_{j}]}{r^{4}}\right\}
\end{aligned}
\end{equation}\\

For $\epsilon_{kl}^{T}=\epsilon_{d}\delta_{kl}$, $\sigma_{kl}^{T}=3K\epsilon_{kl}^{T}=3K\epsilon_{d}\delta_{kl}$, taking this into  \textcolor{blue}{Eq.}\ref{eq2} and noticing the equivalence: $K/G=(2/3)(1+\nu)/(1-2\nu)$, one could get the constrained stress field caused by a pure dilation of eigenstrain in the matrix:

\begin{equation}\label{B3}
\begin{aligned}
&\epsilon_{ij}^{C}=\frac{9(1-2\nu)\epsilon_{d}}{20(1+\nu)(1-\nu)}\left(\frac{a^{5}}{r^{5}}\right)\left(\frac{3x_{i}x_{j}}{r^{2}}\right)+\frac{(1-2\nu)\epsilon_{d}}{4(1+\nu)(1-\nu)}\left(\frac{a^{3}}{r^{3}}\right)\left[(1-2\nu)+\frac{3}{5}\left(\frac{a^{2}}{r^{2}}\right)\right]\\ \\
&\left(\frac{r^{2}\delta_{ij}-3x_{i}x_{j}}{r^{2}}\right)+\frac{3(1-2\nu)\epsilon_{d}}{8(1+\nu)(1-\nu)}\left(1-\frac{a^{2}}{r^{2}}\right)\left(\frac{a^{3}}{r^{3}}\right)\left(\frac{r^{2}\delta_{ij}-8x_{i}x_{j}}{r^{2}}\right)
\end{aligned}
\end{equation}\\
where $\epsilon_{d}=A\sigma^{A}/9K$ for the equivalent eigenstrain due to the uniaxial compressive $\sigma^{A}$.

\section{The pressure analysis acting on the glassy matrix/crystalline interface due to MT}\label{appendixc}

In a Cartesian coordinate system, the traceless strain tensor $\epsilon_{ij}^{T}$ is in generally expressed as\citep{ProcciaPRE}:
\begin{equation}\label{C1}
 \epsilon_{ij}^{T}=(2\lambda_{n}+\lambda_{k})n_{i}n_{j}+(2\lambda_{k}+\lambda_{n})k_{i}k_{j}-(\lambda_{n}+\lambda_{k})\delta_{ij}
  \end{equation}
 where $\lambda_{n}$, $\lambda_{m}$, $\lambda_{k}$,  are the eigenvalues of the strain tensor or the three principal strains with the principal axes aligning along the unit vectors $\mathbf{n}\bot\mathbf{m}\bot\mathbf{k}$. Thus, 
 \begin{align}\label{C2}
&\epsilon_{ij}^{T}g_{i}=\epsilon_{ij}^{T}(x_{i}/r)=[(2\lambda_{n}+\lambda_{k})(\mathbf(n)\centerdot\mathbf{r})n_{j}+(\lambda_{n}+2\lambda_{k})(\mathbf(k)\centerdot\mathbf{r})k_{j}-(\lambda_{n}+\lambda_{k})x_{j}]/r\\ 
&\epsilon_{ij}^{T}g_{i}g_{j}=\epsilon_{ij}^{T}(x_{i}x_{j}/r^{2})=[(2\lambda_{n}+\lambda_{k})(\mathbf(n)\centerdot\mathbf{r})^{2}+(\lambda_{n}+2\lambda_{k})(\mathbf(k)\centerdot\mathbf{r})^{2}-(\lambda_{n}+\lambda_{k})r^{2}]/r^{2}
\end{align}

 When the $\mathbf{m}$ aligns along the direction of $\mathbf{r}$, $\mathbf{n}\centerdot\mathbf{r}=0$, $\mathbf{k}\centerdot\mathbf{r}=0$, $\epsilon_{ij}^{T}g_{i}g_{j}$ takes the maximum negative value of $-(\lambda_{n}+\lambda_{k})$, thus the pressure takes the maximum positive value $p_{max}^{T}=K(\lambda_{n}+\lambda_{k})(1-2\nu)/(1-\nu)$. For the $\lambda_{n}>\lambda_{k}>0>\lambda_{m}=-(\lambda_{n}+\lambda_{k})$, the shear strain size $\epsilon^{T}$ is defined as $\epsilon^{T}=(\epsilon_{1}-\epsilon_{3})/2=(2\lambda_{n}+\lambda_{k})/2$, we have $p_{max}^{T}=K\epsilon^{T}(1+\xi)(1-2\nu)/(1-\nu)$ , where $\xi=\lambda_{n}/(2\lambda_{n}+\lambda_{k})$. When the $\mathbf{n}$ aligns to the vector $\mathbf{r}$, $\mathbf{n}\centerdot\mathbf{r}=1$, $\mathbf{k}\centerdot\mathbf{r}=0$, $\epsilon_{ij}^{T}g_{i}g_{j}$ takes the maximum positive value of $\lambda_{n}$. Thus, the pressure takes the maximum negative value $p_{max}^{C}=-K\epsilon^{T}(1-\psi)(1-2\nu)/(1-\nu)$ , where $\psi=\lambda_{k}/(2\lambda_{n}+\lambda_{k})$.

 \section*{References}



\end{document}